\documentclass[12pt,a4paper]{article}
\pdfoutput=1
\usepackage{graphicx}
\usepackage{soul}
\usepackage[symbol]{footmisc}

\usepackage{amsmath,amsfonts,amssymb}
\usepackage{t1enc}
\usepackage{cite}
\usepackage{colortbl}
\usepackage{pifont}
\usepackage{cancel}

\providecommand{\openone}{\leavevmode\hbox{\small1\kern-3.8pt\normalsize1}}

\begin{document}

\vspace*{-2.5cm}
\begin{flushright}
IFT-UAM/CSIC-19-153 \\
CFTP/20-002
\end{flushright}
\vspace{0.cm}

\begin{center}
\begin{Large}
{\bf Multiphoton signals of a (96 GeV?) stealth boson}
\end{Large}

\vspace{0.5cm}
\renewcommand*{\thefootnote}{\fnsymbol{footnote}}
\setcounter{footnote}{1}
J.~A.~Aguilar--Saavedra$^{a,}$\footnote{On leave of absence from Universidad de Granada, E-18071 Granada, Spain}$^{,}$%
\renewcommand*{\thefootnote}{\arabic{footnote}}\setcounter{footnote}{0}%
\footnote{jaas@ugr.es}, F.~R. Joaquim$^{b,}$\footnote{filipe.joaquim@tecnico.ulisboa.pt} \\[1mm]
\begin{small}
{$^a$ Instituto de F\'isica Te\'orica UAM-CSIC, Campus de Cantoblanco, E-28049 Madrid, Spain} \\ 
{$^b$ Departamento de F\'{\i}sica and CFTP, Instituto Superior T\'ecnico, Universidade de Lisboa, Av. Rovisco Pais 1, 1049-001 Lisboa, Portugal} 
\end{small}
\end{center}

\begin{abstract}
Cascade decays of new scalars into final states with multiple photons and possibly quarks may lead to distinctive experimental signatures at high-energy colliders. Such signals are even more striking if the scalars are highly boosted, as when produced from the decay of a much heavier resonance. We study this type of events within the framework of the minimal stealth boson model, an anomaly-free $\text{U}(1)_{Y'}$ extension of the Standard Model with two complex scalar singlets. It is shown that, while those signals may have cross sections that might render them observable with LHC Run 2 data, they have little experimental coverage. We also establish a connection with a CMS excess observed in searches for new scalars decaying into diphoton final states near 96~GeV. In particular, we conclude that the predicted multiphoton signatures are compatible with such excess. 
\end{abstract}

\renewcommand*{\thefootnote}{\arabic{footnote}}\setcounter{footnote}{0}
\section{Introduction}
The scalar sector of the Standard Model (SM) contains one scalar doublet which spontaneously breaks the gauge symmetry via the Brout-Englert-Higgs mechanism~\cite{Englert:1964et,Higgs:1964ia,Higgs:1964pj}, predicting  the existence of the so-called Higgs boson. A particle compatible with the SM predictions for the Higgs boson and with a mass of approximately 125 GeV was discovered by the ATLAS and CMS Collaborations~\cite{Aad:2012tfa,Chatrchyan:2012xdj} at the Large Hadron Collider (LHC), thereby culminating several decades of searches. Although in the SM the scalar content is the minimal one required to break the symmetry, several additional scalar particles may be present in SM extensions. For instance, models with an augmented gauge symmetry require extra scalar fields to break that symmetry down to the SM group $\text{SU}(3) \times \text{SU}(2)_L \times \text{U}(1)_Y$. Thus, scenarios with heavier scalars decaying into lighter ones (or into pairs of SM weak bosons $V=W,Z$) may be naturally envisaged~\cite{Aguilar-Saavedra:2015iew} as, for example, in the context of left-right models~\cite{Mohapatra:1974gc,Senjanovic:1975rk}. When the former are produced from decays of a very heavy resonance, and the latter decay hadronically, the experimental signature is a multi-pronged fat jet. In this case, the heavier decaying scalar has been dubbed as `stealth boson' for its elusive character~\cite{Aguilar-Saavedra:2017zuc}. 

In ref.~\cite{Aguilar-Saavedra:2019adu} we have proposed  the simplest model that accounts for the said cascade decays --- the minimal stealth boson model (MSBM) --- in which the SM gauge symmetry is enlarged with an extra $\text{U}(1)_{Y'}$ coupling to baryon number up to an arbitrary normalisation constant. The scalar sector comprises two complex SM singlets $\chi_{1,2}$ that, upon $\text{U}(1)_{Y'}$ symmetry breaking, provide masses to the $Z'$ boson and to the new fermions. We point out that, in order for the $Z'$ to decay into scalar pairs, two complex singlets are required, since one of the degrees of freedom is `eaten' by the $Z'$ boson, and also because the required coupling involves one CP-even and one CP-odd scalar weak eigenstate. To ensure gauge-anomaly cancellation, the SM fermion sector is minimally extended with extra leptons.\footnote{This model is denoted by `model 2' in  ref.~\cite{Aguilar-Saavedra:2019adu}; `model 1' is similar, differing only from the fact that vector-like quarks are introduced instead of leptons. The heavy lepton sector has also been considered in refs.~\cite{Duerr:2014wra,Caron:2018yzp,Aguilar-Saavedra:2019iil} in the context of dark-matter phenomenology. Indeed, if $N_1$ is the lightest among the new leptons, then it is a natural dark-matter candidate.}  
In the MSBM one can accommodate sizeable branching ratios (BRs) for the $H_i \to \gamma \gamma$ decays of the new scalars since the corresponding one-loop amplitudes may be enhanced by the Yukawa couplings with the new charged leptons. (We denote the four scalar mass eigenstates as $H_i$, $i=1-4$, being $H_1 \equiv H$ the 125 GeV Higgs boson.) This will be shown in sections~\ref{sec:2} and \ref{sec:3}. After writing down the relevant interactions and the $H_i \to \gamma \gamma$ decay widths in section~\ref{sec:2}, we perform a scan over the parameter space of the model in section~\ref{sec:3} to compute the BRs for $H_i \to \gamma \gamma$, as well as for the other decay modes. We remark that although sizeable BRs for $H_i$ decays into $\gamma \gamma$ are naturally accommodated within the MSBM, they are not a straightforward implication of the model.

Besides the $pp \to H_i \to \gamma \gamma$ signals from direct production of the new scalars, the presence of a heavy $Z'$ resonance opens up the possibility of several conspicuous signals from cascade decays, like those involving collimated photons and/or jets containing hard photons. Their features will be discussed in section~\ref{sec:4} where we will also show that, remarkably, those signals have little experimental coverage, in the sense that the efficiency for such signals in current searches is marginal.
Direct $H_i \to \gamma \gamma$ decays are also interesting by themselves, and may reach detectable levels in some regions of the parameter space. In this regard, it is worth noting that the CMS Collaboration has found~\cite{Sirunyan:2018aui} an excess in the searches for new scalars $h$ decaying into photon pairs with Run 1 (8 TeV) and Run 2 (13 TeV) data, which reaches a statistical significance of 2.8$\sigma$ at $M_h = 95.3$~GeV.
The ATLAS Collaboration has also looked for such process with a larger luminosity of 80 fb$^{-1}$~\cite{ATLAS:2018xad} at 13 TeV, without finding a significant deviation from the background-only expectation at the same mass. However, as noted in ref.~\cite{Heinemeyer:2018wzl}, the limits from the ATLAS Collaboration are weaker and are not able to exclude the CMS excess as due to a new particle. Whether this excess can be due to a stealth boson (dominantly decaying into lighter scalars) will be examined in section~\ref{sec:5}. For completeness, we also explore the possibility that 
the same scalar decays mostly into $b \bar b$. In such case, an excess observed at the Large Electron-Positron (LEP) collider around the same mass~\cite{Barate:2003sz} can also be accommodated. This possibility is further explored in appendix~\ref{sec:a}. To conclude, in section~\ref{sec:6}  we discuss our results in light of current and future experimental searches.

\section{$H_j \rightarrow \gamma \gamma $ decays in the MSBM}
\label{sec:2}

The MSBM~\cite{Aguilar-Saavedra:2019adu} extends the SM scalar content with two complex scalar fields $\chi_1$ and $\chi_2$, which are SM singlets and are equally charged under a gauged U(1)$_{Y'}$ symmetry. The anomaly-cancellation conditions require extra matter fields which, in this work, we consider to be a set of two vector-like lepton singlets $N'_{1}$, $E'_{1}$ and a doublet $(N'_{2},E'_{2})$ (we use primes on fermion weak eigenstates in order to distinguish them from unprimed mass eigenstates). The SM and U(1)$_{Y'}$ hypercharges $Y$ and $Y^\prime$ of the various fields are shown in table~\ref{tab:Ynew}, where $Q_{iL}$ and $l_{iL}$ denote the left-handed (LH) quark and lepton $\text{SU}(2)_L$ doublets of the SM, respectively. The right-handed (RH) quarks and lepton singlets are $u_{iR}$, $d_{iR}$ and $e_{iR}$.
A model with the same fermion content and hypercharge assignments as those in table~\ref{tab:Ynew}, but with only one scalar singlet, has been previously considered~\cite{Duerr:2014wra,Caron:2018yzp}. However, the decays of the $Z'$ boson into two scalars require two or more singlets, opening also the possibility of pure scalar cascade decays, and implying a richer phenomenology.
\begin{table}[htb]
\begin{center}
\begin{tabular}{ccccccc}
& $Y$ & $Y'$ & & & $Y$ & $Y'$ \\
\hline
$Q_{Li}=(u'_{iL}\;d'_{iL})^T$ & $1/6$ & $Y'_q$ & \quad & $u'_{iR}$  & $2/3$  & $Y'_q$ \\
                 &           &        &           & $d'_{iR}$  & $-1/3$ & $z$ \\ 
$l_{iL}=(\nu'_{iL}\;e'_{iL})^T$ & $-1/2$ & $0$ & \quad & $e'_{iR}$  & $-1$  & $0$ \\
$\Phi=(\phi^+\;\phi^0)^T$ &$1/2$ & 0\\
\hline
$N'_{1L}$       & $0$     & $9 Y'_q/2$ & \quad & $N'_{1R}$       & $0$ &  $-9Y'_q/2$  \\
$E'_{1L}$       & $-1$    & $9 Y'_q/2$ & \quad & $E'_{1R}$       & $-1$ &  $-9Y'_q/2$\\
$\ell_L=(N'_{2L}\;E'_{2L})^T$ & $-1/2$ & $-9 Y'_q/2$ & \quad & $\ell_R=(N'_{2R}\;E'_{2R})^T$ & $-1/2$ &  $9Y'_q/2$  \\
$\chi_1$       & $0$    & $9 Y'_q$ & \quad & $\chi_2$       & $0$ &  $9Y'_q$\\
\hline
\end{tabular}
\end{center}
\caption{SM and $\text{U}(1)_{Y'}$ hypercharge assignments ($Y$ and $Y^\prime$, respectively) for the SM fields and the vector-like leptons, with $Y'_q$ a free parameter.}
\label{tab:Ynew}
\end{table}

The most general scalar potential invariant under SU(2)$_L\times$U(1)$_Y \times$U(1)$_{Y'}$ is $V = V_{Z_2} + V_{\not Z_2}$, with
\begin{eqnarray}
V_{Z_2} & = & m_{0}^2 \Phi^\dagger \Phi + m_{11}^2 \chi_1^\dagger \chi_1 + m_{22}^2 \chi_2^\dagger \chi_2 
\notag \\
& &+ \frac{\lambda_0}{2} (\Phi^\dagger \Phi)^2  + \frac{\lambda_1}{2} (\chi_1^\dagger \chi_1)^2
 + \frac{\lambda_2}{2} (\chi_2^\dagger \chi_2)^2
+ \lambda_3 (\chi_1^\dagger \chi_1) (\chi_2^\dagger \chi_2) \notag \\
& & + \frac{1}{2} \left[  \lambda_4 (\chi_1^\dagger \chi_2) (\chi_1^\dagger \chi_2) + \text{h.c.} \right]
+ \frac{\lambda_5}{2} (\Phi^\dagger \Phi) (\chi_1^\dagger \chi_1)
+ \frac{\lambda_6}{2} (\Phi^\dagger \Phi) (\chi_2^\dagger \chi_2)
\,, \notag \\
V_{\not Z_2} & = &  m_{12}^2  \chi_1^\dagger \chi_2 +\frac{1}{2} \left[  \lambda_7 (\chi_1^\dagger \chi_2) (\chi_1^\dagger \chi_1)
+ \lambda_8 (\chi_1^\dagger \chi_2) (\chi_2^\dagger \chi_2) 
+ \lambda_9 (\Phi^\dagger \Phi)  (\chi_1^\dagger \chi_2)\right] + \text{h.c.}  \,,
\label{ec:VS}
\end{eqnarray}
where $\Phi=(\phi^+\;\phi^0)$ is the SM Higgs doublet. The terms in $V_{Z_2}$ ($V_{\not Z_2}$) conserve (break) a $Z_2$ symmetry under which only $\chi_2$ transforms non trivially as $\chi_2\rightarrow - \chi_2$. While $m_0^2$, $m_{11}^2$, $m_{22}^2$, $\lambda_{0-3}$ and $\lambda_{5,6}$ are real, $m_{12}^2$, $\lambda_{4}$ and $\lambda_{7-9}$ can be, in general, complex. We define
\begin{align}
& \phi^0 = \frac{1}{\sqrt 2} (\rho_0+v + i \eta_0 ) \,, \quad \chi_1 = \frac{1}{\sqrt 2} (\rho_1+u_1 + i \eta_1 ) \,, \quad
\chi_2 = \frac{1}{\sqrt 2} (\rho_2 + i \eta_2+ u_2 e^{i \varphi}) \,, 
\end{align}
where $\rho_i$ and $\eta_i$ are real fields and
\begin{align}
& \langle \phi^0 \rangle = \frac{v}{\sqrt 2}\,, \quad \langle \chi_1 \rangle = \frac{u_1}{\sqrt 2}  \,, \quad
\langle \chi_2 \rangle = \frac{u_2 \,e^{i \varphi}}{\sqrt 2} \,,
\label{eq:vevs}
\end{align}
with $v = 246\,{\rm GeV}$. As usual, we will also use
\begin{align}
u=\sqrt{u_1^2 + u_2^2}\;\;,\;\;\tan\beta=\frac{u_2}{u_1}\,.
\label{eq:tbu}
\end{align}
A detailed analysis of the scalar potential and scalar mass spectrum can be found in ref.~\cite{Aguilar-Saavedra:2019adu}. Upon U(1)$_{Y'}$ symmetry breaking, the neutral $Z'$ gauge boson acquires the mass
\begin{equation}
M_{Z'}^2 = (g_{Z'} Y'_{\chi})^2 \,u^2 \,,
\label{eq:MZpdef}
\end{equation}
where $g_{Z'}$ and $Y'_{\chi}$ are the U(1)$_{Y'}$ coupling constant and $\chi_{1,2}$ hypercharge, respectively. All $Z'$-scalar interactions can be found in ref.~\cite{Aguilar-Saavedra:2019adu}. 

There are two would-be Goldstone bosons in the model, namely $G_1^0 = \eta_0$ and $G_2^0 = \cos \beta \, \eta_1 + \sin \beta \, \eta_2$. The orthogonal state $A^0 = - \sin \beta \, \eta_1 + \cos \beta \, \eta_2$ is CP-odd, being a mass eigenstate in case the scalar potential parameters are real and $\varphi = 0$. In general, $H'_i = (\rho_0 \; \rho_1 \; \rho_2 \; A^0)$ are related to the mass eigenstates $H_i = (H_1 \; H_2 \; H_3 \; H_4)$ by
\begin{align}
H'_i=O_{ij}H_j\,,
\label{eq:Hrot}
\end{align}
where $O$ is an orthogonal $4\times 4$ real matrix parameterised in terms of $2\times 2$ rotations as $O = \widehat{O}_{34} \widehat{O}_{24} \widehat{O}_{14} \widehat{O}_{23} \widehat{O}_{13} \widehat{O}_{12}$, where $\widehat{O}_{kl}$ corresponds to a rotation in the $(k,l)$ plane by an angle $\theta_{kl}$. We identify the 125 GeV SM Higgs boson $H$ as being $H_1$. The scalar interactions with the $Z'$ boson field originate from the term
\begin{equation}
\mathcal{L} = i g_{Z'} Y'_\chi \left(\chi_1^* \overleftrightarrow{\partial_\mu} \chi_1 + \chi_2^* \overleftrightarrow{\partial_\mu} \chi_2 \right) B^{\prime \mu} \,,
\end{equation}
which in the mass-eigenstate basis reads
\begin{equation}
\mathcal{L}_{Z' H_i H_j} = g_{Z'} Y'_{\chi} R_{ij} H_i \overleftrightarrow{\partial_\mu} H_j \, Z^{\prime \mu} \,,
\end{equation}
with $i < j$ and mixing factors
\begin{equation}
R_{ij} = \cos \beta \left[ O_{4i} O_{3j} - O_{4j} O_{3i} \right] - \sin \beta  \left[ O_{4i} O_{2j} - O_{4j} O_{2i} \right] \,.
\label{ec:Rij}
\end{equation}
Notice that $R_{ij}$ are anti-symmetric and therefore $R_{ii} = 0$, reflecting the fact that $Z' \to H_i H_i$ is forbidden. Also, it can be shown that $\sum_{i<j} R_{ij}^2 = 1$ due to the orthogonality of the mixing matrix $O$.

\subsection{Scalar-fermion interactions in the mass basis}

Given their relevance for the computation of scalar diphoton decay amplitudes, we now obtain the scalar-fermion interactions in the mass basis. The masses of the new fermions and their interactions with scalars are determined by the gauge-invariant Lagrangian
%
%
%
%
\begin{align}
-\mathcal{L}_{\rm Y} & = 
( y_{1}^E \chi_1 + x_{1}^E \chi_2 ) \overline{E'_{1L}} E'_{1R}
+ (y_{1}^N \chi_1 + x_{1}^N \chi_2 ) \overline{N'_{1L}} N'_{1R}
+ (y_2 \chi_1^\ast + x_2\chi_2^\ast) \,\overline{\ell_L} \ell_R \nonumber  \\
& + \overline{\ell_L} ( w_{1}^E\Phi E'_{1R} + w_{1}^N \widetilde{\Phi} N'_{1R}) + (w_{2}^E \overline{E'_{1L}} \Phi^\dag + w_{2}^N \overline{N'_{1L}} \widetilde{\Phi}^\dag) \ell_R + {\rm h.c.}\,,
\label{eq:LY}
\end{align}
where the Yukawa couplings $x_i$, $y_{ i}^F$ and $w_{i}^F$ (with $i=1,2$, $F=E,N$) are general complex numbers. Taking into account eqs.~(\ref{eq:vevs}), the corresponding mass terms for the new charged and neutral fermions in the interaction basis $F'_{L,R}=(F'_1\;F'_2)_{L,R}^T$ are
\begin{equation}
\mathcal{L}_F = - \overline{F'_L} M_F F'_R + {\rm h.c.} \,,
\end{equation}
with
\begin{equation}
M_F=\frac{1}{\sqrt{2}}
\left(
\begin{array}{cc}
y_{1}^F u_1+x_{1}^F u_2 e^{i\varphi}      & w_{2}^F v  \\
w_{1}^F v     & y_2 u_1+x_2 u_2 e^{-i\varphi} 
\end{array}
\right)
\,.
\label{eq:MFdef}
\end{equation}
We define the field rotations to the mass-eigenstate basis $F_{L,R}=(F_1\;F_2)_{L,R}^T$ as
\begin{equation}
F'_{L,R}=U_{L,R}^F F_{L,R} \,,\quad
U_L^{F \dagger}\,M_F\, U_R^F = {\rm diag}(m_{F_1},m_{F_2}) \,.
\label{eq:psidef}
\end{equation}
Here, $U_{L,R}^F$ are $2\times 2$ unitary complex matrices and the masses of the new leptons $m_{F_i}$ are real and positive. To compute $U_{L,R}^F$, one defines the Hermitian matrices $H_L^F=M_F M_F^{\dag}$ and $H_R^F=M_F^{\dag }M_F$ diagonalised by $V_{L,R}^{F\dagger}\,H_{L,R}^F\, V_{L,R}^F={\rm diag}(m_{F_1}^2,m_{F_2}^2)$, being $V_{L,R}^F$ unitary matrices given by:
\begin{align}
V_{L,R}^F=\left(
\begin{array}{cc}
\cos\theta_{L,R}^F      & e^{i\delta_{L,R}^F}\sin\theta_{L,R}^F \\
-e^{-i\delta_{L,R}^F} \sin\theta_{L,R}^F     & \cos\theta_{L,R}^F
\end{array}
\right)\,.
\label{eq:VLVRdef}
\end{align}
The mixing angles $\theta_{L,R}^F$ and CP phases $\delta_{L,R}^F$ satisfy
\begin{align}
\tan(2\theta_{L,R}^F)=\frac{2\left|(H_{L,R}^F)_{12}\right|}{(H_{L,R}^F)_{22}-(H_{L,R}^F)_{11}} \,, \quad
\delta_{L,R}^F=\arg(H_{L,R}^F)_{12}\,.
\label{eq:tLtRdef}
\end{align}
Since, in general, $V_{L,R}^F$ diagonalise $M_F$ up to diagonal phases, an additional phase transformation must be performed in either the LH or RH fields to express the interactions in the physical mass basis. Namely, we follow the convention
\begin{align}
& U_L^F = V_L^F \,, \notag \\
& U_R^F = V_R^F \cdot {\rm diag}(e^{-i\delta_1},e^{-i\delta_2}) \,,\quad
\delta_{j}={\rm arg}(V_L^{F\dagger} M_F V_R^F)_{jj} \,.
\label{eq:ULURdef}
\end{align}
Considering eqs.~\eqref{eq:Hrot}, \eqref{eq:LY} and \eqref{eq:psidef}, we can write the scalar-fermion interactions in the mass-eigenstate basis as:
\begin{equation}
-\mathcal{L}_{HFF}=(A_{jka}^F+iB_{jka}^F)\,\overline{F_{jL}} F_{kR} H_a + {\rm h.c.} \,, 
\label{eq:LHpsi2}
\end{equation}
with $(j,k)=1,2$, $a=1,\dots,4$, and the coefficients
\begin{align}
A_{jka}^F & = \frac{O_{1a}}{\sqrt{2}}\sum_{m\neq n=1,2}
(U_L^F)_{mk} (U_R^F)^\ast_{nj} w_n^F
+\frac{1}{\sqrt{2}} \sum_{m=1,2}
(U_L^F)_{mk} (U_R^F)^\ast_{mj} (O_{3a} x_m^F +O_{2a} y_m^F) \,,\nonumber\\
B_{jka}^F & = \frac{O_{4a}}{\sqrt{2}}\sum_{m=1,2} (-1)^m
(U_L^F)_{mk} (U_R^F)^\ast_{mj} (y_m^F \sin\beta - x_m^F \cos\beta) \,, \label{eq:ABdef}
\end{align}
with $y_2^F \equiv y_2$, $x_2^F \equiv x_2$.
In terms of $F = F_L+F_R$, eq.~\eqref{eq:LHpsi2} reads
\begin{align}
-\mathcal{L}_{HFF} = \overline{F_{j}} (C_{jka}^F + i D_{jka}^F \gamma_5) F_{k} H_a \,,
\label{eq:LHpsi3}
\end{align}
again with $(j,k)=1,2$, $a=1,\dots,4$ and the coefficients $C_{jka}^F$, $D_{jka}^F$ given by:
\begin{align}
C_{jka}^F & = \frac{1}{2}\left[A_{jka}^F + A_{kja}^{F \ast} + i (B_{jka}^F - B_{kja}^{F \ast}) \right]\,,\nonumber\\
D_{jka}^F & = \frac{1}{2}\left[A_{jka}^F - A_{kja}^{F \ast} + i (B_{jka}^F + B_{kja}^{F \ast}) \right]\,.
\label{eq:CDdef}
\end{align}
\subsection{$H_j \rightarrow \gamma \gamma$ decay widths}

\begin{figure}[t]
\begin{center}
\begin{tabular}{ccc}
\includegraphics[height=3.1cm,clip=]{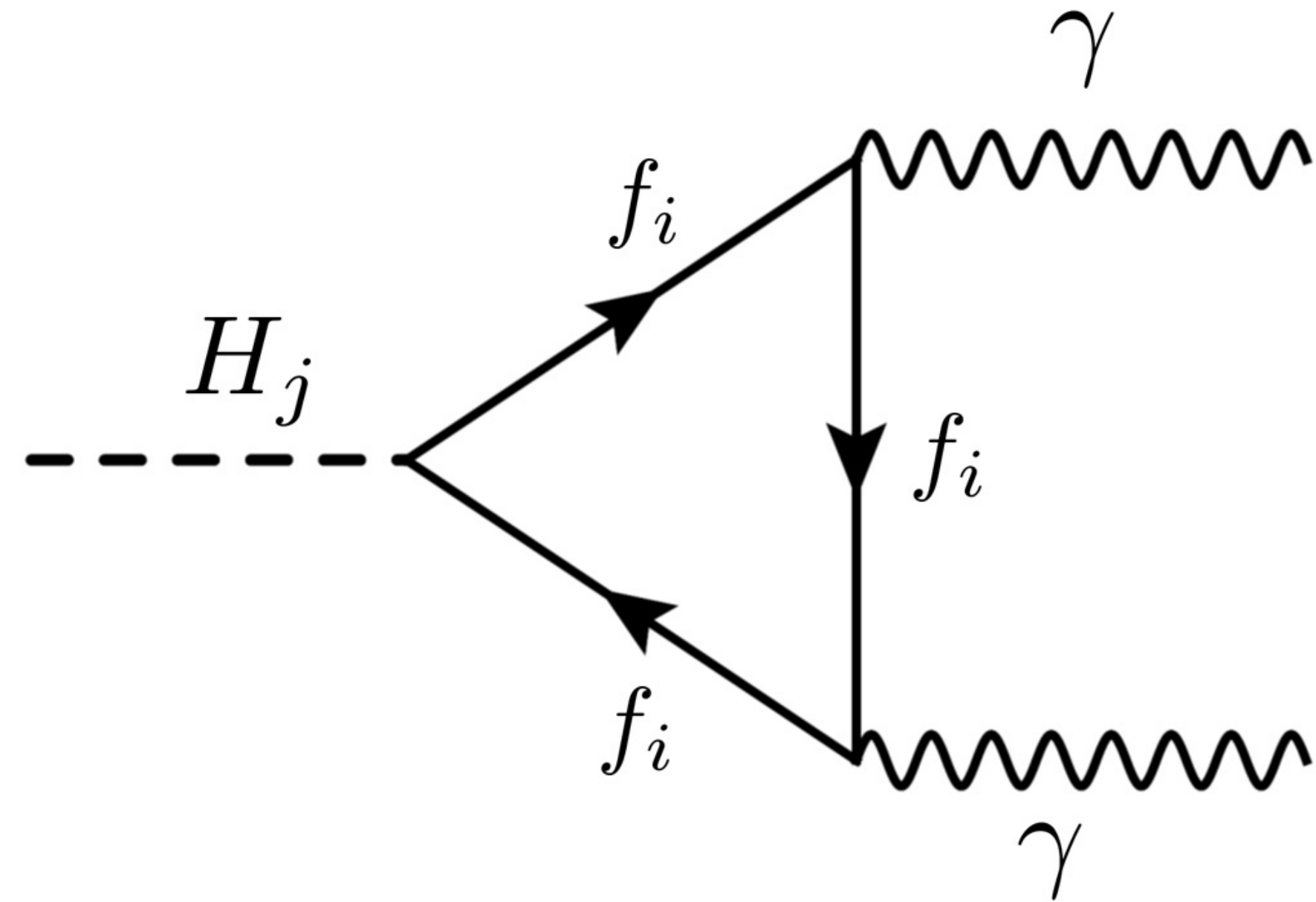} &
\includegraphics[height=3.1cm,clip=]{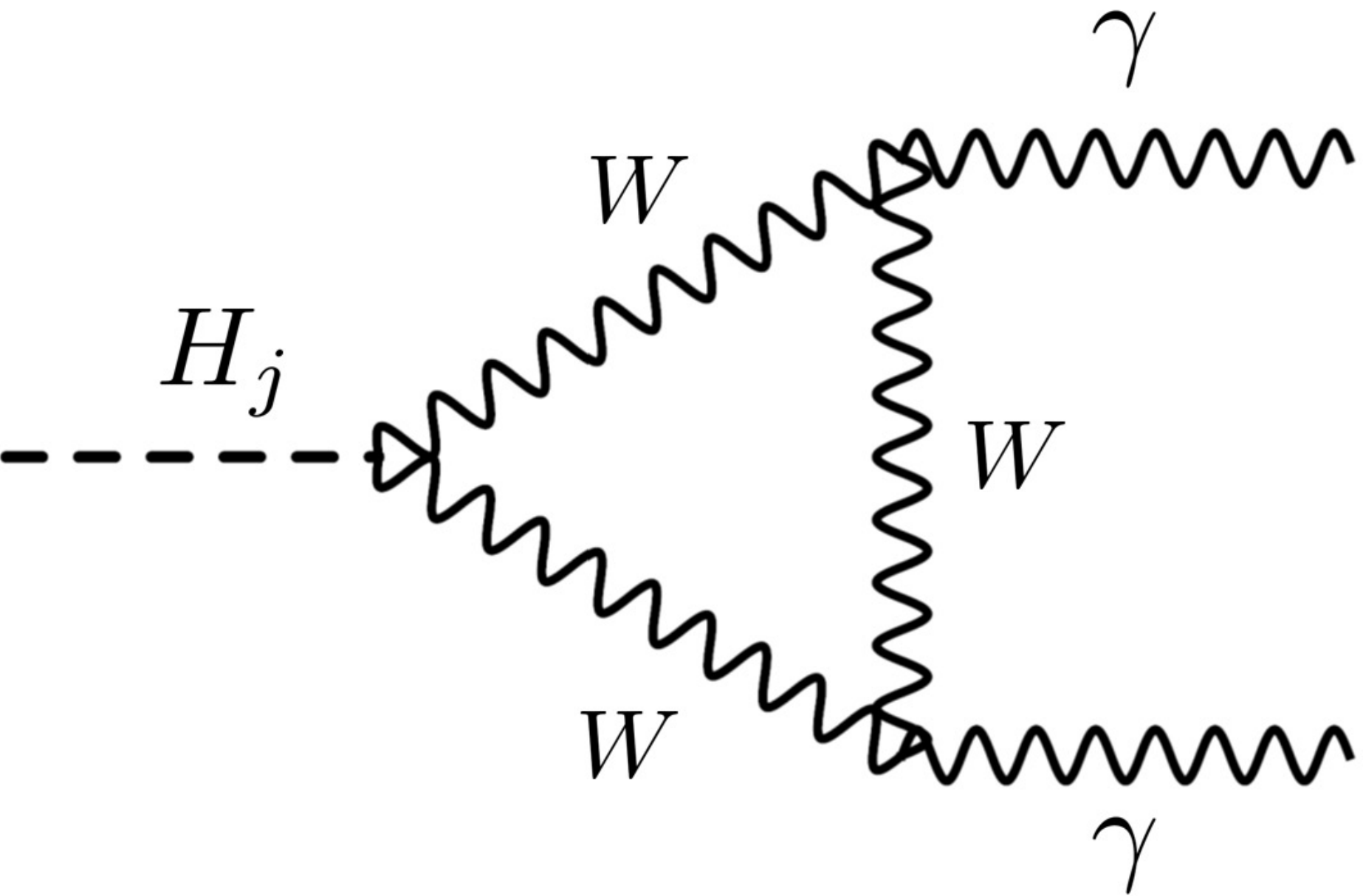} & 
\includegraphics[height=3.1cm,clip=]{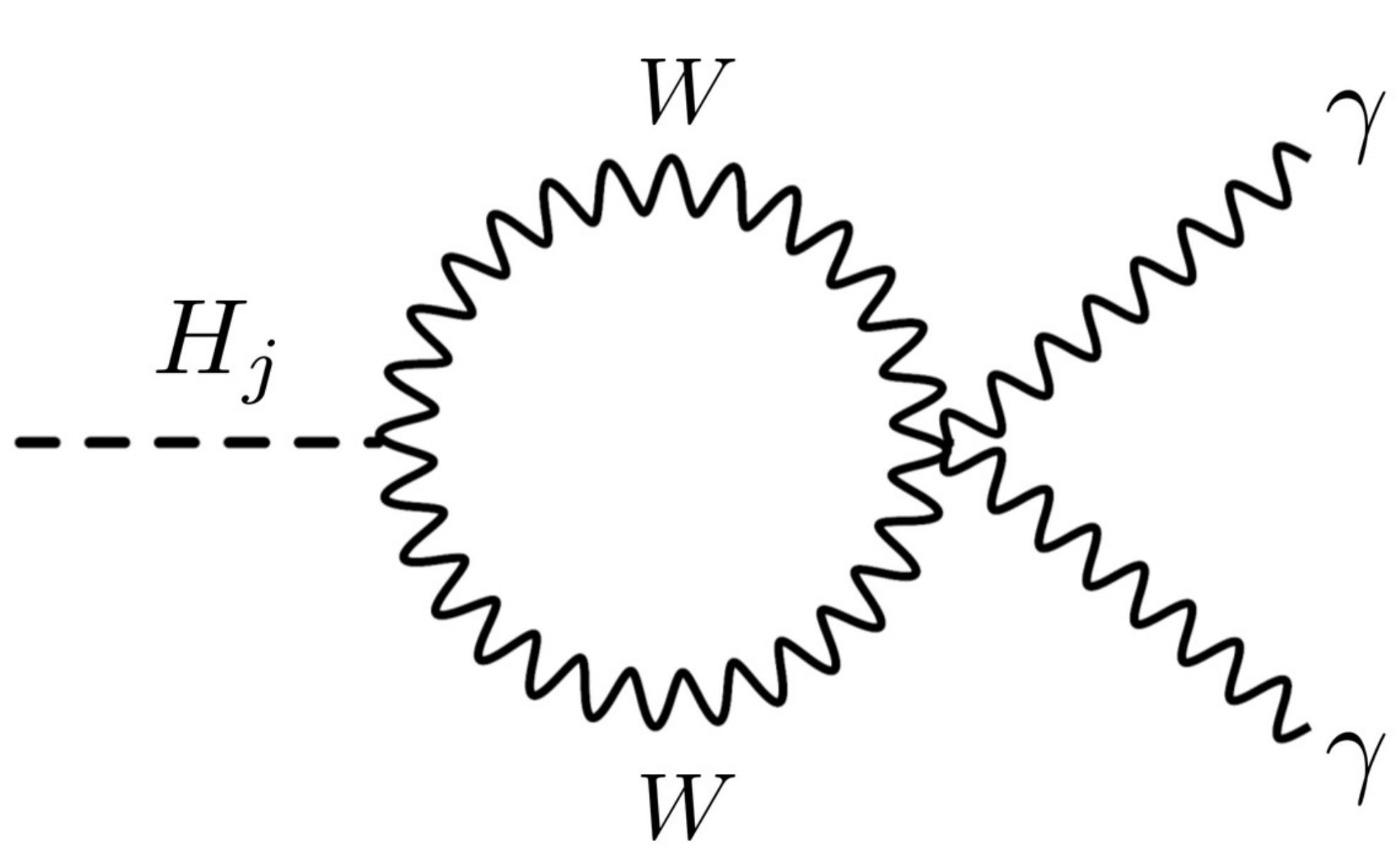} \\
(a) & (b) & (c)
\end{tabular}
\caption{One-loop diagrams contributing to $H_j\rightarrow \gamma \gamma$. In the fermion loop (a) $f$ stands for any electrically-charged fermion (see table~\ref{tab:Ynew}).}
\label{fig:Hgg}
\end{center}
\end{figure}

The Feynman diagrams which contribute at one-loop level to $H_j \rightarrow \gamma \gamma$ decays are shown in figure~\ref{fig:Hgg}.
In diagram (a) $f$ is any electrically-charged fermion in the model, namely $f=e_i,u_i,d_i,E_i$. Due to scalar mixing, all charged fermions enter the $H_j$ diphoton decay loop, although in practice some of the contributions are suppressed due to experimental constraints on the SM Higgs couplings. Diagrams (b) and (c) stand for the $W$-boson contributions. It is convenient to write the $\bar{f} f H_j$ interactions of eq.~\eqref{eq:LHpsi3} in the form:
\begin{equation}
-\mathcal{L}_{Hff}=\frac{m_{f}}{v}(a_j^{f}+i\,b_j^{f}\gamma_5)\bar f f H_j \,,
\label{eq:LHff2}
\end{equation}
with $j=1,\dots,4$. The scalar and pseudoscalar couplings $a_f$ and $b_f$, respectively, are given by
\begin{equation}
a_j^{f} = O_{1j} \,,\quad b_j^{f}=0 \,,
\label{ec:abSM}
\end{equation}
for SM fermions, where $O$ is the scalar mixing matrix defined in eq.~\eqref{eq:Hrot}. Notice that $O_{1j}$ is the admixture between the $\text{SU}(2)_L$ doublet $\Phi$ and the $j$-th scalar eigenstate; the SM fermions do not couple to the $\text{SU}(2)_L$ singlets. For the new leptons,
\begin{eqnarray}
a_j^{E_k} =\dfrac{v}{m_{E_k}} C_{kkj}^E\,,\quad b_j^{E_k}=\dfrac{v}{m_{E_k}} D_{kkj}^E \,,
\label{eq:abcoef}
\end{eqnarray}
being the coefficients $C_{kkj}^E$ and $D_{kkj}^E$ those of \eqref{eq:CDdef}. As for the $WWH_j$ coupling, we have:
\begin{equation}
\mathcal{L}_{HWW}=g\,O_{1j}m_W W_\mu^+W^{\mu-}H_j\,,
\label{eq:LWWH}
\end{equation}
where $g$ is the SM SU(2) gauge coupling and $m_W$ the $W$ boson mass.

Due to electromagnetic gauge invariance, the tensor form of the $H_j\rightarrow \gamma \gamma$ amplitude can be generically expressed as~\cite{Ilisie:2014hea}
\begin{equation}
\mathcal{M}_{H_j\gamma\gamma}=\mathcal{M}_j^{\mu\nu}\varepsilon^\ast_\mu(q) \varepsilon_\nu^\ast(p) \,,\quad 
\mathcal{M}^{\mu\nu}_j=(g^{\mu\nu}p \cdot q-p^\mu q^\nu)S_j^X+\epsilon^{\mu\nu\alpha\beta}p_\alpha q_\beta\,\tilde{S}_j^X \,,
\label{eq:AmpHgg}
\end{equation}
being $p$ and $q$ the momenta of the outgoing photons, while $\varepsilon(p)$ and $\varepsilon(q)$ denote the corresponding polarisation vectors. The quantities $S_j^X$ and $\tilde{S}_j^X$ with $X=f,W$, are scalar form factors computed by considering the relevant contributions. In the present case, these stem from the one-loop diagrams shown in figure~\ref{fig:Hgg}, for which
\begin{align}
& S_j^f =-\frac{\alpha}{\pi v}\ \sum_{f}a_j^{f} Q_{f}^2 N_c^{f} \left[
\tau_{f}+(\tau_{f}-1)g(\tau_{f})\right]\tau_{f}^{-2}\,, \notag \\
& S^W_j=\frac{\alpha}{2\pi v}O_{1j}\,\tau_W^{-2}[\,3(2\tau_W-1)g(\tau_W)+2\tau_W^{2}+3\tau_W] \,, \notag \\
& \tilde{S}_j^f =-\frac{\alpha}{\pi v} \sum_{f} b_j^{f} Q_{f}^2 N_c^{f} g(\tau_{f}) \tau_{f}^{-1}\,, \notag \\
& \tilde{S}^W_j =0\,.
\label{eq:Ffactors}
\end{align}
Here, $\alpha=e^2/(4\pi)$ is the electromagnetic fine-structure constant, $Q_{f}$ is the fermion electric charge, $N_c^{f}$ is the number of colours of $f$ and
$\tau_X=M_{H_j}^2 / 4m_X^2$.
The function $g(\tau)$ is given by~\cite{Gunion:1989we}
\begin{align}
g(\tau)=\left\{ 
\begin{array}{ll}
\arcsin^2{\sqrt{\tau}} &{\rm if}\,\tau \leq 1\\
-\dfrac{1}{4}
\left[\ln\left(
\dfrac{1+\sqrt{1+\tau^{-1}}}{1+\sqrt{1-\tau^{-1}}}\right)-i\pi\right] &{\rm if}\,\tau > 1\\
\end{array}\right.\,.
\label{eq:gtau}
\end{align}
As usual, the decay widths are obtained using
\begin{equation}
\Gamma(H_j\rightarrow \gamma \gamma)=\frac{|\vec{p}\,|}{8\pi M_{H_j}^2}\overline{|\mathcal{M}_{H_j\gamma\gamma}|^2}=\frac{G_F\alpha^2 M_{H_j}^3}{128\sqrt{2}\pi^3}
\left(
|S_j^f+S_j^W|^2+|\tilde{S}_j^f|^2
\right)\,,
\end{equation}
where $|\vec{p}\,|=M_{H_j}/2$ and a $1/2$ factor has been included for identical particles in the final state.

\section{Parameter-space analysis}
\label{sec:3}

In order to investigate the possible multiphoton signals from $H_j$ decays in the MSBM, we proceed with a parameter-space analysis. For simplicity, we restrict ourselves to the case with $\lambda_{7-9}=0$ in eq.~\eqref{ec:VS}, which corresponds to having the $Z_2$ symmetry softly broken by the term $m_{12}^2\neq 0$. The interested reader is addressed to ref.~\cite{Aguilar-Saavedra:2019adu} for details on the full analysis of the scalar potential, and on the reconstruction of its parameters in terms of the scalar masses $M_{H_i}$ and mixings $O_{ij}$. Here, we briefly summarise the general procedure. We will consider the following values for scalar masses and for the $Z'$ mass and coupling:
\begin{align}
& M_{Z'}=2.2\,{\rm TeV} \,,\quad g_{Z'} Y_q'=0.15 \,, \notag \\
& M_{H_1}=125~{\rm GeV} \,,\quad M_{H_3}=96~{\rm GeV} \,,\quad M_{H_4}=105~{\rm GeV} \,.
\label{eq:input}
\end{align}
The TeV-scale mass for $Z'$ boson is chosen so as to have the scalars from its decay highly boosted. For better comparison with previous work~\cite{Aguilar-Saavedra:2019adu} we choose the value of 2.2~TeV. The chosen coupling $g_{Z'} Y_q'$ fulfills the limits from dijet production (see next section), while $M_{H_3}$ is set to 96 GeV in order to establish a connection with section~\ref{sec:5}, where we investigate if a CMS excess can be due to one of the scalars present in the spectrum. (Of course, the signals studied in section~\ref{sec:4} may take place for a wider range of masses.) We choose $M_{H_4}$ of the same order. The mass of the lightest scalar $H_2$ is an output parameter which in our scan we require to be within the interval $[20,40]$~GeV in order to allow $H_3 \to H_2 H_2$ and $H_4 \to H_2 H_2$. The values of the six angles $\theta_{ij}$ defining the scalar mixing matrix $O$ in eq.~\eqref{eq:Hrot}, and of the scalar potential parameter $\lambda_2$, are randomly varied in the intervals $[0,2\pi]$ and $[-1,1]$, respectively. The latter range is enough to obtain the predictions later presented.

These inputs allow us to determine $\tan\beta$, $m_{H_2}$, $\lambda_{0-6}$ and $m_{12}^2$. As done in ref.~\cite{Aguilar-Saavedra:2019adu}, we will only keep those points which lead to a global minimum of the potential with $v,u_{1,2}\neq 0$ and $\varphi=0$. Notice also that, since we are considering the case with a softly broken $Z_2$ symmetry under which $\chi_2 \rightarrow -\chi_2$, the couplings $x_2$ and $x_1^{N,E}$ in eq.~\eqref{eq:LY} vanish. We have verified that this simplification does not affect the predictions for the $H_i$ decays, as there is still enough freedom for the coefficients in eqs.~(\ref{eq:ABdef}).
Fermion mixing plays little role in our study and, given that $v\ll u \simeq 1.6\,{\rm TeV}$, we will neglect it in eq.~\eqref{eq:MFdef} implying $\theta_{L,R}^{F}=0$ and, thus, $U_{L,R}^F=\openone_{2\times 2}$. The only relevant couplings among the new fermions and the scalars are then $y_1^{E,N}$ and $y_2$, which we randomly vary in the perturbative interval $[0,4\pi]$. The resulting heavy leptons masses range from 300 GeV to 10 TeV. The $\lambda$ couplings are small because the four scalars are relatively light, $m_{H_i} \ll u$, which guarantees perturbativity.

Experimental constraints on the signal-strength parameters for each production and decay mode combination of the SM Higgs $H$ must be taken into account. Since there are no new coloured particles in our model, the SM Higgs production cross sections for the various processes ($gg$ fusion, vector boson fusion (VBF), $VH$ associated production, $t \bar t H$, etc.) are all rescaled by the mixing factor $O_{11}^2\leq 1$, implying
\begin{equation}
\mu_{\gamma \gamma} \equiv \frac{\sigma(pp \to H \to \gamma\gamma )} {\sigma(pp \to H \to \gamma \gamma)_{\rm SM}} =
 O_{11}^2 \frac{\text{BR}(H \to \gamma \gamma)}{\text{BR}(H \rightarrow \gamma \gamma)_{\rm SM}} \,,
\label{eq:mugg}
\end{equation}
for any of those processes. Here, ${\rm BR}(H\rightarrow \gamma \gamma)$ is computed considering the results obtained in the previous section for the diphoton Higgs decay in the MSBM. The subindex `SM' obviously refers to the quantities within the SM. In our scan we use the naive weighted average (without including possible correlations) of the $\mu_{\gamma \gamma}$ values corresponding to the different Higgs production processes, reported by the ATLAS collaboration in ref.~\cite{Aad:2019mbh},
\begin{equation}
\mu_{\gamma\gamma}=1.025\pm 0.121\,.
\label{eq:muggval}
\end{equation}
We use as constraint the agreement of the model prediction with this value within one standard deviation.

For the remaining SM decay modes of the Higgs boson, which we generically denote with the shorthand $H \to \cancel{\gamma\gamma}$ (with $\cancel{\gamma \gamma} = ZZ,WW,bb,\tau\tau,gg$, no sum over channels),
the widths are suppressed by the factor $O_{11}^2$. Then, we have
\begin{equation}
\mu_{ \cancel{\gamma\gamma} } \equiv \frac{\sigma(pp \to H \to \cancel{\gamma\gamma})} {\sigma(pp \to H \to \cancel{\gamma\gamma})_{\rm SM}} =
\frac{O_{11}^4 \Gamma_{\rm SM}}{O_{11}^2 \sum \Gamma(H\rightarrow \cancel{\gamma\gamma})_{\rm SM}+\Gamma(H\rightarrow \gamma\gamma)+ \Gamma_{\rm new}}\,,
\label{eq:muHiggs}
\end{equation}
where $\Gamma_\text{SM}$ is the SM Higgs total width, $\Gamma_\text{new}$ the width into new modes (e.g. lighter scalars) and the sum in the denominator comprises all SM decay channels other than $\gamma \gamma$. Note that, as defined above, $\mu_{ \cancel{\gamma\gamma} }$ is the same for all the individual channels and production processes. Computing the naive weighted average of the $\mu_{ \cancel{\gamma\gamma} }$ signal strengths for non-diphoton decays given in ref.~\cite{Aad:2019mbh}, one obtains $\mu_{\cancel{\gamma\gamma}}=1.070 \pm 0.096$. Restricting ourselves to the $\mu_{\cancel{\gamma\gamma}} \leq 1$ interval that has physical meaning within this model, the constraint translates into the $1\sigma$ lower limit
\begin{equation}
\mu_{ \cancel{\gamma\gamma} }\geq 0.931\,,
\label{eq:murestval}
\end{equation}
which we require in our scan. In addition, we include in our scan the limit
\begin{equation}
O_{12}^2 \leq 0.02 \,,
\end{equation}
from direct searches for new scalars at LEP~\cite{Barate:2003sz}. The limits on $H_4$ from LEP are weaker than the indirect constraints from Higgs measurements. On the other hand, there are limits on its decay into $\gamma \gamma$. Using $\text{BR}(H_4 \to \gamma \gamma)_\text{SM} = 1.77 \times 10^{-3}$~\cite{Heinemeyer:2013tqa}, the limits in ref.~\cite{Sirunyan:2018aui} translate into the constraint
\begin{equation}
O_{14}^2 \, \text{BR}(H_4 \to \gamma \gamma) \leq 5.3 \times 10^{-4} \,.
\end{equation}
Direct limits from the production of $H_{3,4}$ and cascade decay $H_{3,4} \to H_2 H_2 \to 4b$ were reviewed in ref.~\cite{Aguilar-Saavedra:2019adu} (see also section~\ref{sec:6}). In section~\ref{sec:4} we mention, when relevant, the limits from processes involving the decays $H_2 \to \gamma \gamma$. 

Indirect limits from the $S$ and $T$ parameters~\cite{Peskin:1990zt,Peskin:1991sw} do not lead to further constraints in the parameter space. The contribution of the new scalars is quite small because it is suppressed by the small mixings $O_{1j}^2$, and the new scalars have masses that are not far away from $M_H$. Using the expressions in ref.~\cite{Aguilar-Saavedra:2019ghg}, we find that the extra contribution to $T$ is at the level of $2 \times 10^{-3}$, and the extra contribution to $S$ is of the order of $8 \times 10^{-3}$, in good agreement with the latest determinations (assuming $U=0$) of $T = 0.06 \pm 0.06$, $S = 0.02 \pm 0.07$~\cite{Tanabashi:2018oca}. The contributions of the lepton singlets to $S$ and $T$ vanish, and so does the contribution of the vector-like lepton doublet in the limit considered of no mixing, since in that case the two mass eigenstates are degenerate (see ref.~\cite{Tanabashi:2018oca} for a review). 

The purely scalar contributions to electric dipole moments vanish because the pseudoscalar component of the coupling to SM fermions vanishes, see eq.~(\ref{ec:abSM}). The only contributions arise from two-loop Barr-Zee~\cite{Barr:1990vd} diagrams with closed loops of heavy leptons and exchange of a scalar $H_j$ and a $Z$ boson or photon. The amplitudes are proportional to
\begin{equation}
a_j^{E_k} b_j^{E_k} = (-1)^k O_{1j} O_{4j} \tan \beta \,,
\end{equation}
where we have simplified eqs.~(\ref{eq:ABdef}) for the case of vanishing heavy lepton mixing. Therefore, the individual amplitudes are already suppressed by small mixing factors, and 
in most of the parameter space fulfiling the rest of contstraints $\tan \beta \leq 1$.
Moreover, the contributions of the two heavy leptons have opposite sign and there is a (partial) cancellation between them, which is exact when $m_{E_1} = m_{E_2}$. Furthermore, there is a partial cancellation between the contributions of the several scalars, since $\sum_{j=1}^4 O_{1j} O_{4j} = 0$ by unitarity. As a result, the contributions to electric dipole moments are below experimental bounds, as we have explicitly verified for the case of the electron.

\begin{figure}[t]
\begin{center}
\begin{tabular}{ccc}
\includegraphics[height=5.5cm,clip=]{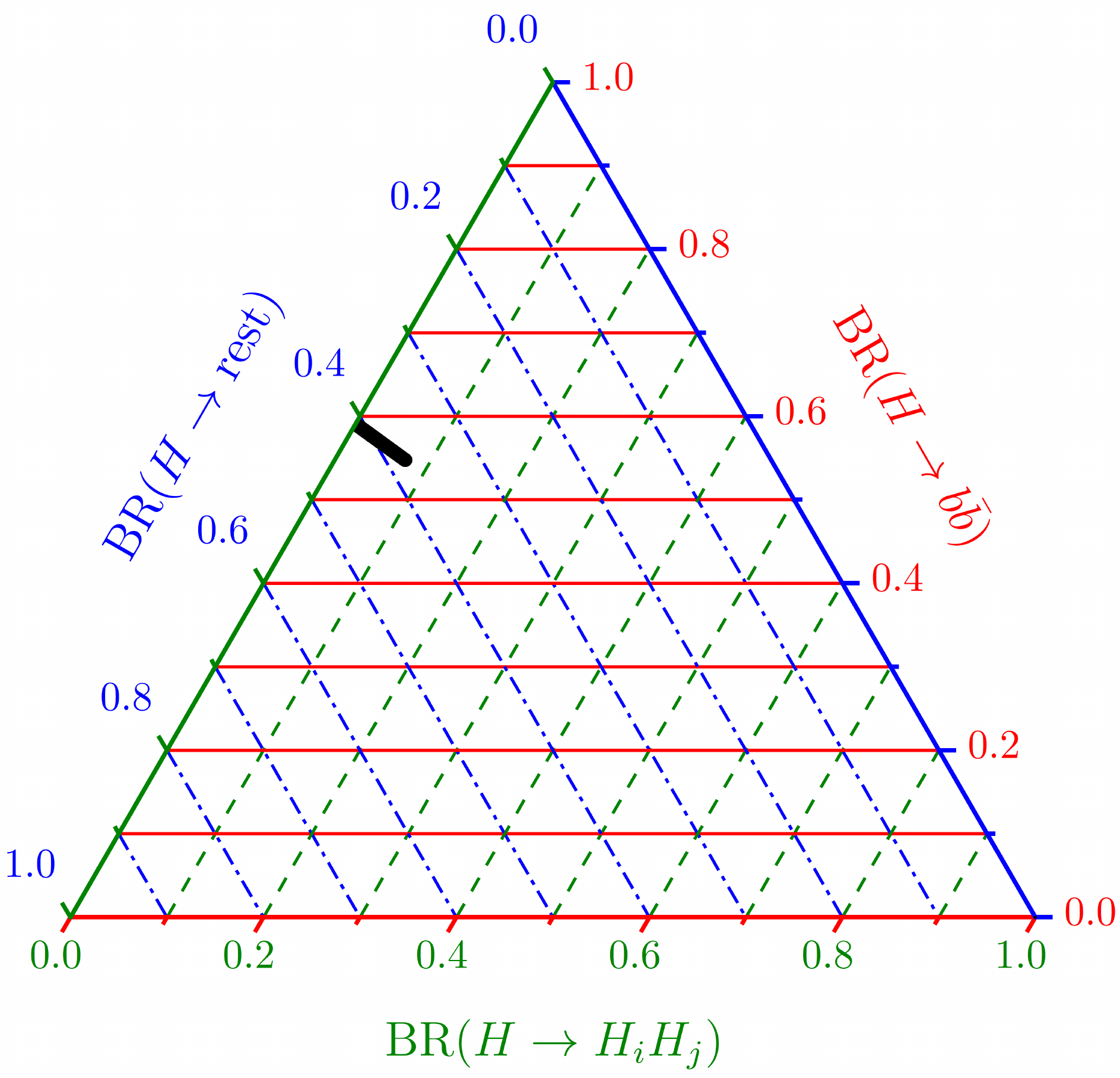} & \quad &
\includegraphics[height=5.5cm,clip=]{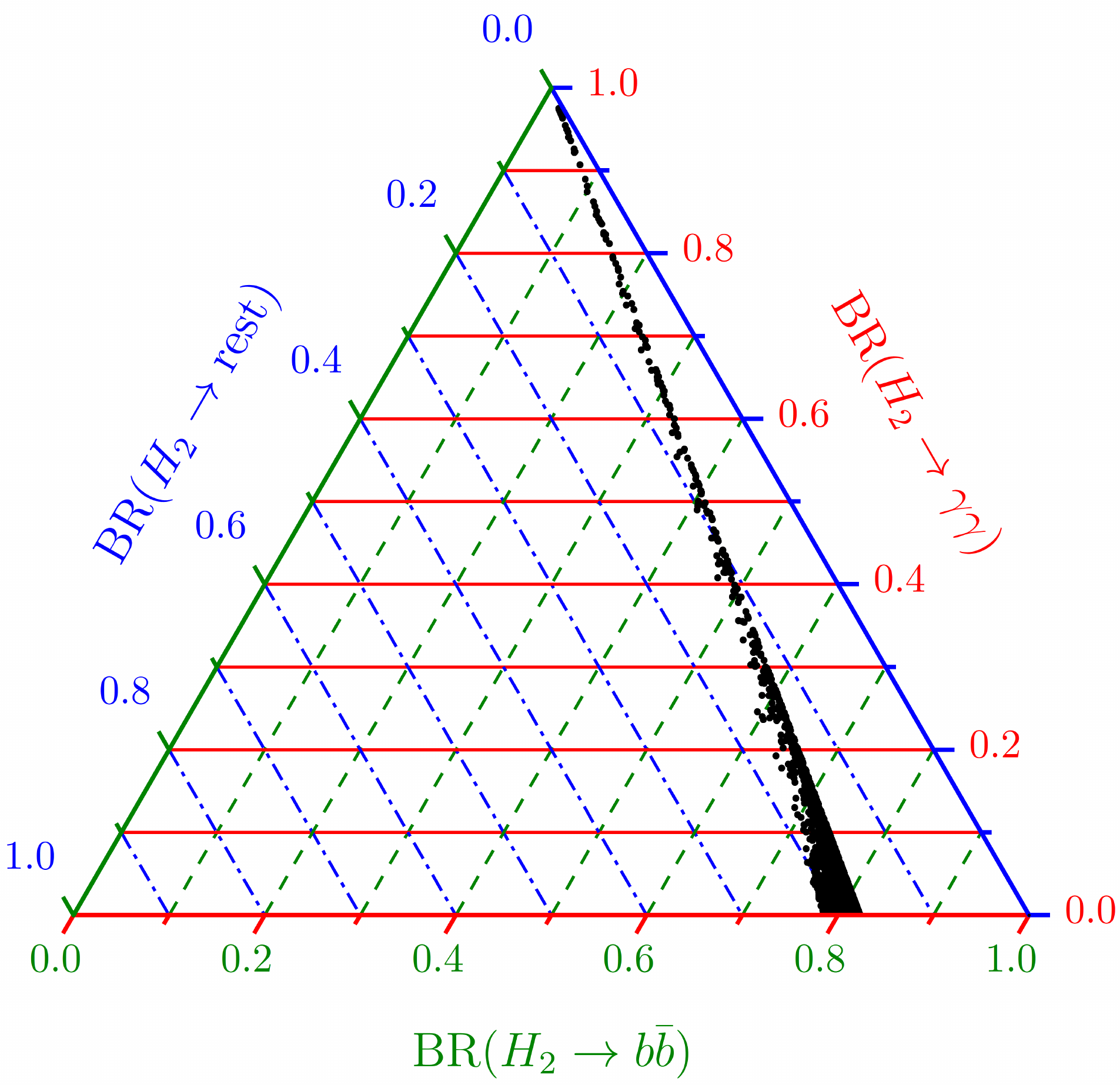} \\[2mm]
\includegraphics[height=5.5cm,clip=]{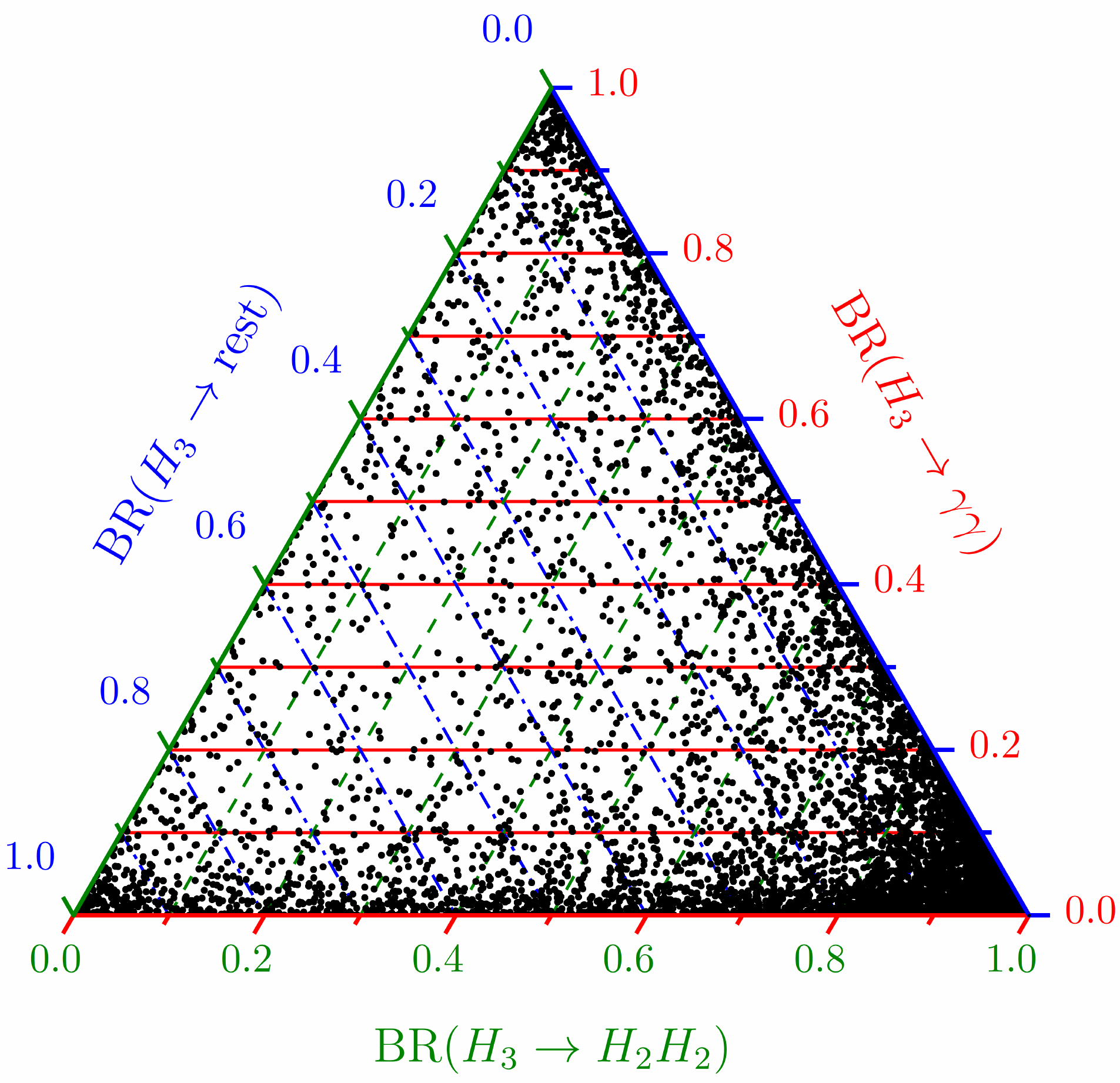} & &
\includegraphics[height=5.5cm,clip=]{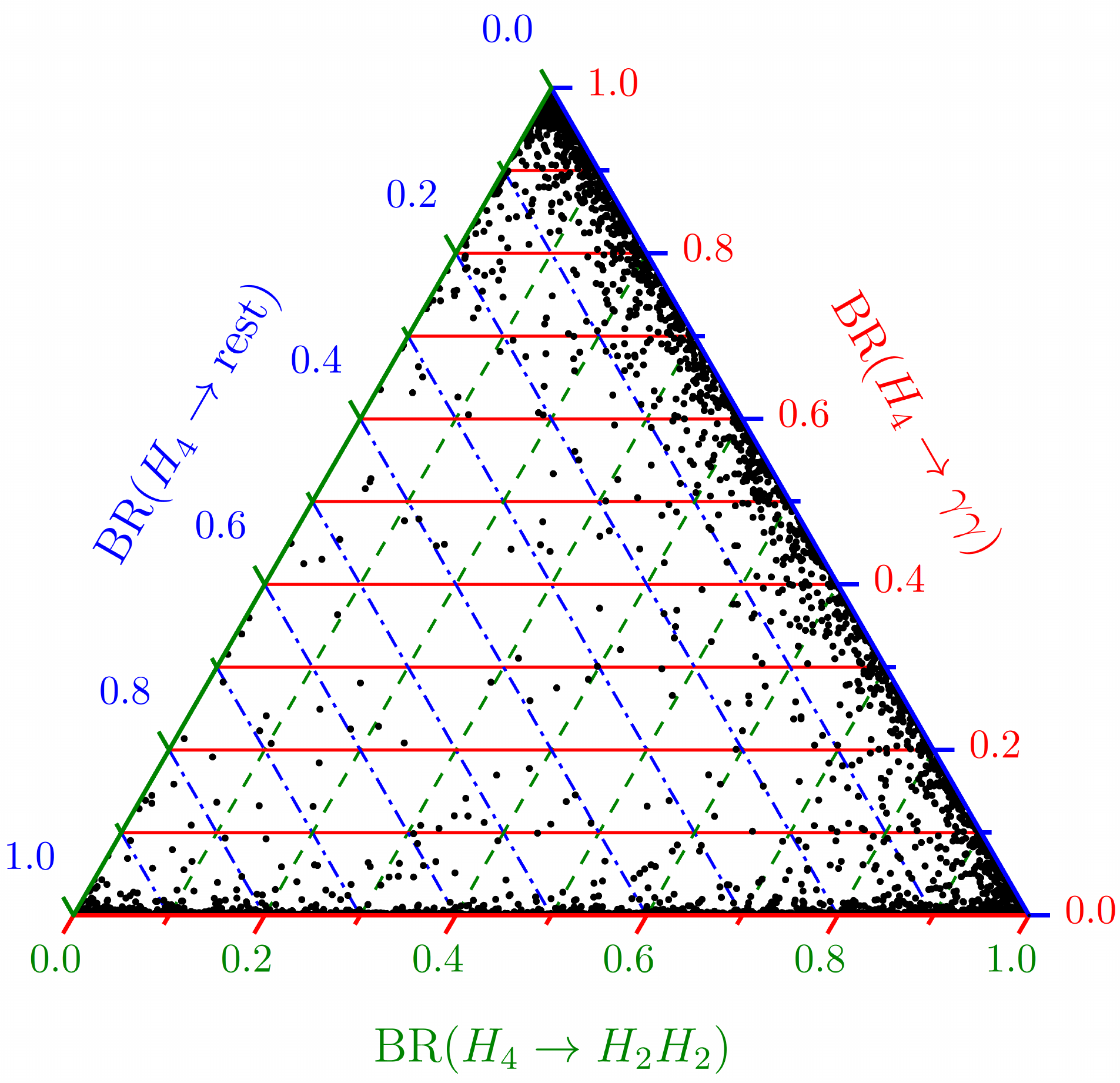}
\end{tabular}
\caption{Decay BRs for the four $H_i$ scalars obtained from a scan over parameter space with the inputs \eqref{eq:input}. All points obey the constraints on the Higgs signal-strength parameters in eqs.~\eqref{eq:muggval} and \eqref{eq:murestval}. In each plot, the label `rest' refers to the channels not shown in the other axes.}
\label{fig:Hdecays}
\end{center}
\end{figure}

In figure~\ref{fig:Hdecays} we show in ternary plots the results obtained for the decay BRs of the four scalars, corresponding to $6 \times 10^4$ allowed points in parameter space. As one can see from the top-left panel of this figure, the aforementioned constraints on the Higgs signal-strength parameters imply for the SM-like Higgs boson ${\rm BR}(H\rightarrow H_i H_j) \lesssim 0.1$ with $0.55 \lesssim {\rm BR}(H\rightarrow b \bar{b}) \lesssim 0.6$ (the SM value is approximately 0.58~\cite{deFlorian:2016spz}). As expected, the lightest scalar $H_2$ predominantly decays into $\gamma\gamma$ and $b\bar{b}$ (top-right panel), being the probability to decay into other modes ($\tau \tau$, $c\bar c$ and $gg$, labeled as `rest') around 20$\%$, at most. Notice that $\text{BR}(H_2\rightarrow\gamma \gamma)$ can even reach unity. For the remaining two scalars $H_{3,4}$ (bottom panels) the trend is less clear, though they are expected to predominantly decay into $H_2 H_2$ and $\gamma \gamma$. Notice that $\text{BR}(H_{3,4}  \to \gamma \gamma) = 1$ (apex of the triangle) is possible, as well as $\text{BR}(H_{3,4} \to H_2 H_2) = 1$ (lower-right vertex).

\begin{figure}[t!]
\begin{center}
\begin{tabular}{ccc}
\includegraphics[height=5.5cm,clip=]{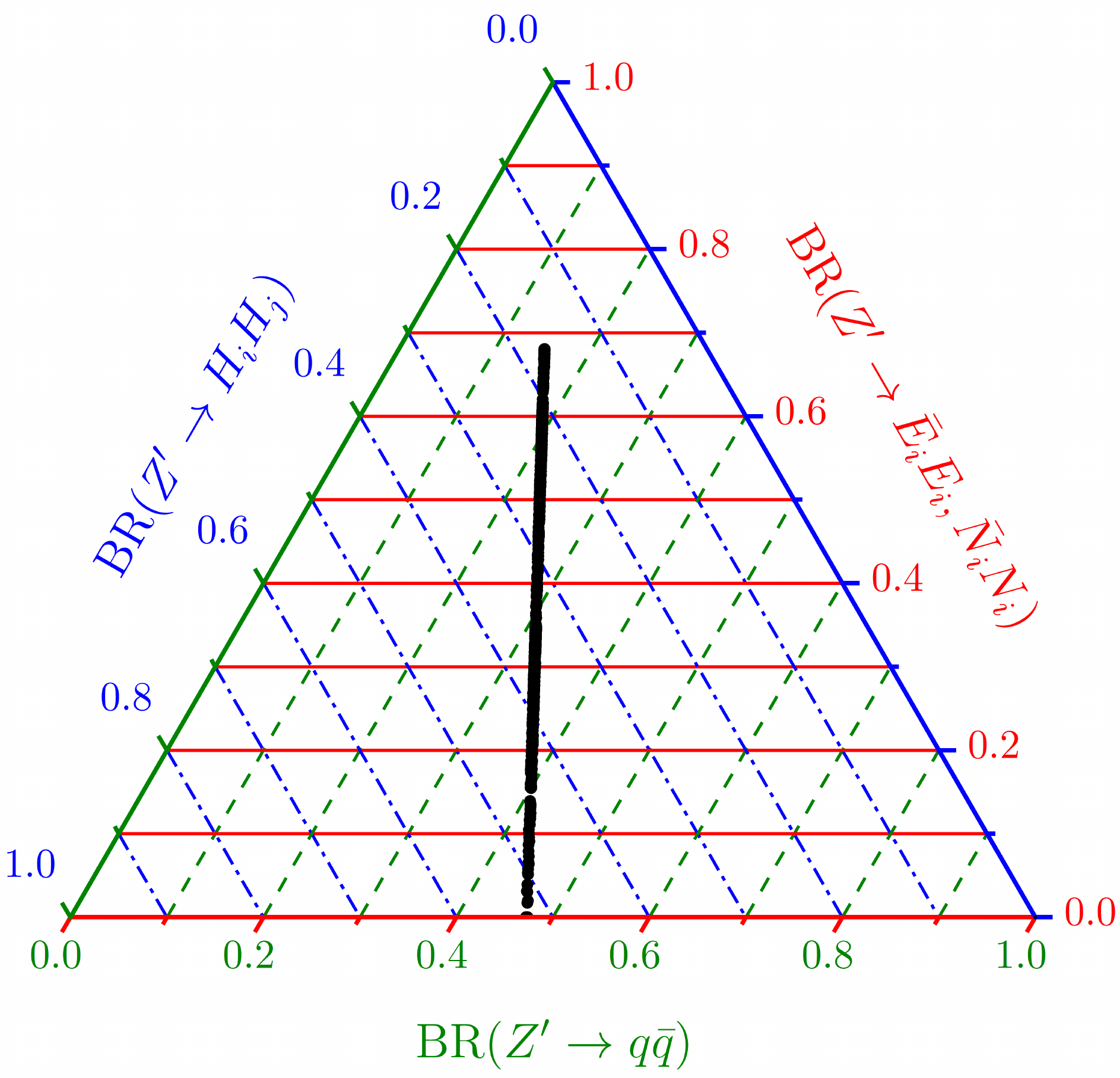} & \quad
\includegraphics[height=5.5cm,clip=]{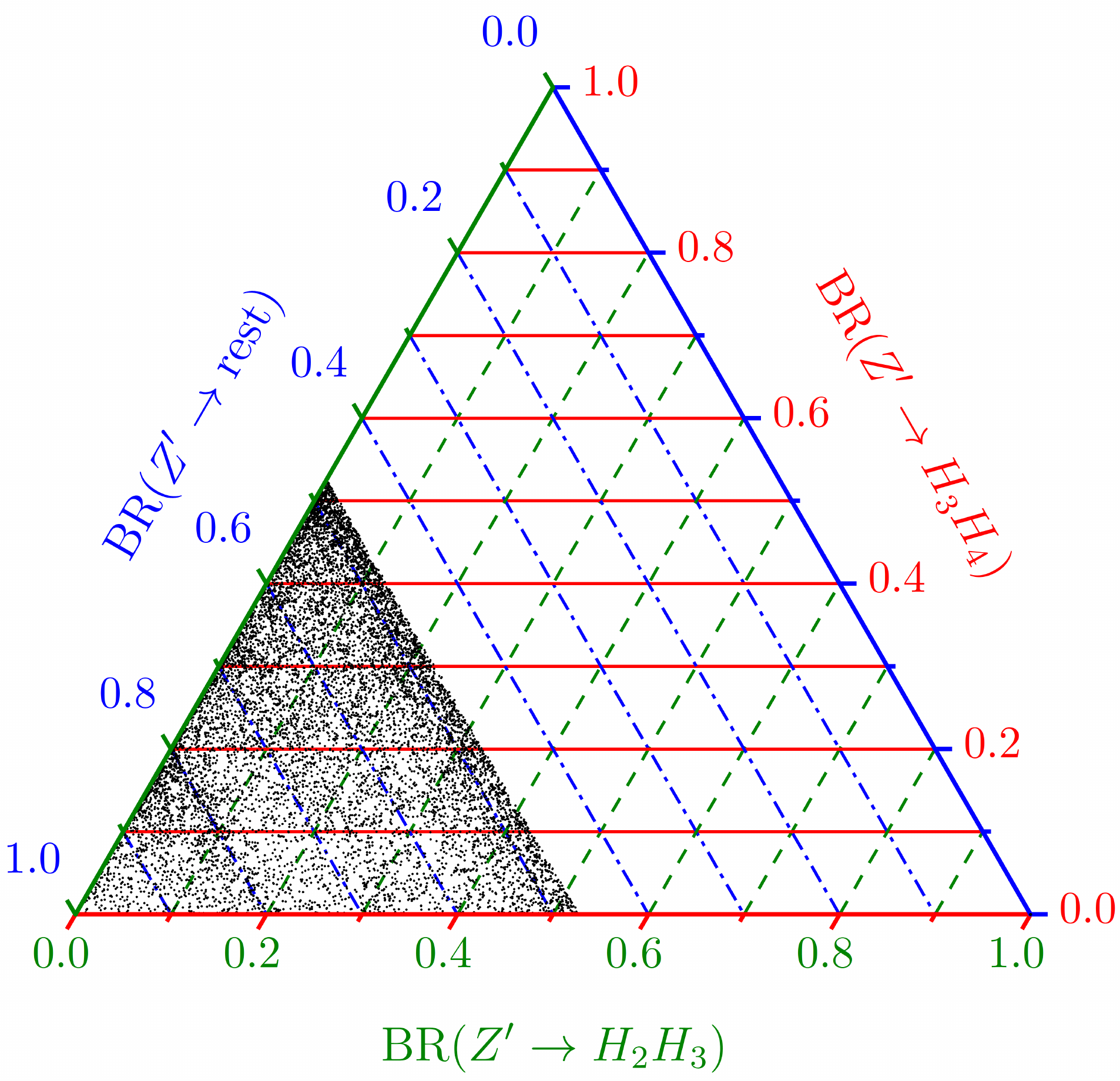} 
\end{tabular}
\caption{Decay BRs for the $Z'$ boson  considering the same points shown in figure~\ref{fig:Hdecays}. As in that figure, all points obey the constraints on the Higgs signal-strength parameters in eqs.~\eqref{eq:muggval} and \eqref{eq:murestval}.}
\label{fig:Zpdecays}
\end{center}
\end{figure}

The decay BRs of the $Z'$ boson presented in figure~\ref{fig:Zpdecays} show little dependence on scalar mixing. In the left panel the decays into pairs of quarks, heavy leptons and scalars are compared. The relative BRs basically depend on the masses of the new leptons. When these are heavy, the corresponding $Z'$ decay modes are kinematically forbidden (base of the triangle), while if $m_F \ll M_{Z'}$ they dominate the $Z'$ decays with $\text{Br}(Z' \to F \bar F) = 36/53 \simeq 0.68$. In the right panel we show $\text{BR}(Z' \to H_3 H_4)$ and $\text{BR}(Z' \to H_2 H_3)$, which are the most interesting ones for the discussion in the next section. (The decays involving the SM Higgs $H$ are very suppressed by mixing, and $Z' \to H_2 H_4$ follows a similar pattern as $Z' \to H_2 H_3$.) As it can be seen from the same plot, these decays can have a BR up to 50\%. All these results are in agreement  with those in ref.~\cite{Aguilar-Saavedra:2019adu} for model 2. 

In summary, figures~\ref{fig:Hdecays} and \ref{fig:Zpdecays} reflect the fact that there is a variety of interesting signals coming from the cascade decays of the heavy resonance $Z'$ into scalar pairs, allowed by present constraints. The scenario corresponding to $Z' \to H_3 H_4$, with $H_{3,4} \to H_2 H_2$ followed by $H_2 \to b \bar b$ provides the perfect template for the stealth-boson hypothesis, as investigated in refs.~\cite{Aguilar-Saavedra:2017zuc,Aguilar-Saavedra:2019adu}. On the other hand, final states with multi-pronged jets and/or multiphotons have been overlooked in the literature. In the following section we will discuss some interesting features of these signals in the context of the MSBM.

\section{Multiphoton signals}
\label{sec:4}

When scalars $H_i$ are produced either directly or from the decay of a heavy resonance, their decays may lead to multiphoton signals. We will review their main features in this section, with results obtained at the parton level. For the signal generation we use {\scshape MadGraph5}~\cite{Alwall:2014hca} with the model implemented using {\scshape Feynrules}~\cite{Alloul:2013bka} and the universal Feynrules output~\cite{Degrande:2011ua}. We do not aim at a sensitivity estimate for these signals, as their detection relies on tools (e.g. the identification of collimated photons and photons within jets) that are not available at the level of fast simulation.

The possible cross section for these signals is determined in first place by the  $Z'$ production cross section which is
\begin{equation}
\sigma = 63.4 \left( \frac{g_{Z'} Y'_q}{0.1} \right)^2  ~\text{fb}\,
\end{equation}
for the $M_{Z'} = 2.2$ TeV benchmark point considered. For this mass, dijet constraints~\cite{Aad:2019hjw} set an upper limit $\sigma(pp \to Z' \to jj) \times A \leq 24.3$ fb at the 95\% CL, where $A$ is the acceptance for the event selection in the rapidity region $|y| \leq 0.6$. We evaluate the acceptance by simulating a $Z' \to jj$ sample using {\scshape Pythia}~\cite{Sjostrand:2007gs} and {\scshape Delphes}~\cite{deFavereau:2013fsa}, with jets reconstructed using {\scshape FastJet}~\cite{Cacciari:2011ma}. The resulting limits on the coupling depend on the masses of the heavy leptons $F=E,N$, and range between
\begin{align}
& |g_{Z'} Y'_q| \leq 0.19 && m_{F} > M_{Z'}/2 \,, \notag \\
& |g_{Z'} Y'_q| \leq 0.33 && m_{F} \ll M_{Z'} \,.
\label{ec:limg}
\end{align}
Note however that the ATLAS limits are obtained under the assumption of a $Z'$ width $\Gamma_{Z'} \leq 0.15 M_{Z'}$, which is not always the case. Therefore, the actual limits may be looser. 
For $M_{H_i} \ll M_{Z'}$ (which is the case in our benchmark), and depending on $m_F$, the $Z' \to H_i H_j$ branching ratios are
\begin{align}
m_{F} > M_{Z'}/2:\quad& \sum_{i<j} \text{BR}(Z' \to H_i H_j) = 9/17\,, \notag \\ 
m_{F} \ll M_{Z'}:\quad& \sum_{i<j} \text{BR}(Z' \to H_i H_j) = 9/53 \,. 
\label{ec:BRH}
\end{align}
Since the SM scalar doublet $\Phi$ does not couple to the $Z'$ boson, the decay modes involving the Higgs boson $H$ turn out to be very suppressed by mixing. The decays into the remaining scalar pairs $H_2 H_3$, $H_2 H_4$ and $H_3 H_4$ may be comparable, or the decay into one pair may dominate over the others, as seen in figure~\ref{fig:Zpdecays}. If we assume for definiteness $m_{F} > M_{Z'}/2$, we then have 
\begin{equation}
\sigma (pp \to Z'  \to H_i H_j) = 33.5   \left( \frac{g_{Z'} Y'_q}{0.1} \right)^2  ~\text{fb}\,,
\label{ec:xsecmax2}
\end{equation}
for the signals we study in the remainder of this section, with a maximum of 121 fb.

The first signal addressed, boosted diphotons, is produced from $H_{2-4} \to \gamma \gamma$, when the  scalars are produced from $Z'$ decay. Instead, the second and third signals arise from the (expected) dominant decay $H_{3,4} \to H_2 H_2$, when one or the two lighter scalars decay into photons. These signals are possible if the lightest scalar has a sizeable branching ratio into $\gamma\gamma$, as it may be the case in our benchmark scenario (see figure~\ref{fig:Hdecays}). Note that in the remainder of this section we discuss the type of `object' produced by one of the boosted scalars in $Z' \to H_i H_j$; an actual event will have two of such objects, or one plus a hadronically-decaying scalar, etc.

\subsection{Boosted diphotons from $H_i \to \gamma \gamma$}
\label{sec:6.1}

The scalars $H_i$ can be directly produced, and their decay into two photons gives signals such at the CMS excess investigated in the next section. On the other hand, when $H_i$ originate from the decay of a heavy resonance (in our case the $Z'$ boson), they produce two photons that in the laboratory frame are relatively close in pseudo-rapidity $\eta$ and azimuthal angle $\phi$. These diphotons usually fail to pass the isolation criteria for prompt photons imposed in experimental analyses. To clarify this point, we briefly review how photons are identified.

Photons are reconstructed from energy depositions in the electromagnetic calorimeters (ECAL) of the ATLAS and CMS detectors. (The discrimination between photons and electrons is based on tracks in the inner detectors.) In the central region, the calorimeter cells have a size $\Delta \eta \times \Delta \phi = 0.025 \times 0.0245$ for the ATLAS ECAL~\cite{Aaboud:2016yuq}\footnote{The ATLAS ECAL has a first layer with a higher granularity, this size refers to the second layer where most of the photon energy is deposited.} and $0.0174 \times 0.0174$ for the CMS ECAL~\cite{Khachatryan:2015iwa}. Photons are reconstructed from $3 \times 5$ clusters in ATLAS, yielding a cluster size of  $\Delta \eta \times \Delta \phi = 0.075 \times 0.124$. In CMS $5 \times 5$ clusters are used, with a size $\Delta \eta \times \Delta \phi = 0.087 \times 0.087$. In both experiments, additional rows up to $\Delta \phi = 0.172$ (ATLAS) and  $\Delta \phi = 0.609$ (CMS) may be added in the case of photon conversion into an $e^+ e^-$ pair, which are curved in opposite directions in $\phi$ due to the magnetic field. For the photons so identified, there are several `reference' sets of photon isolation conditions, from tighter to looser, which require, among other, the absence of additional radiation above some given thresholds in a cone of $\Delta R \equiv \sqrt{(\Delta \eta)^2 + (\Delta \phi)^2} \leq 0.2-0.3$~\cite{Khachatryan:2015iwa,Aaboud:2016yuq}. 

\begin{figure}[t]
\begin{center}
\begin{tabular}{cc}
\includegraphics[height=5cm,clip=]{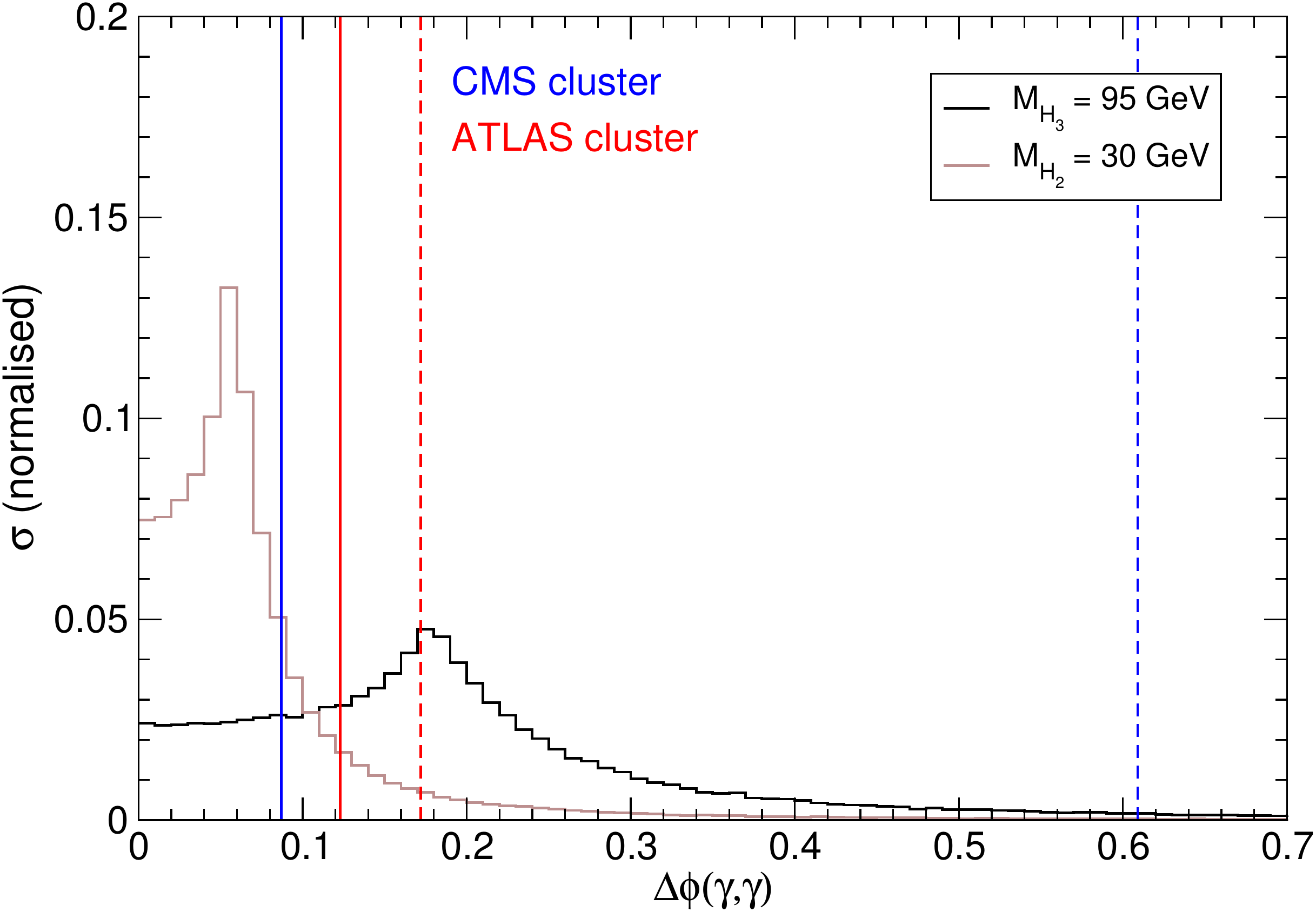} & 
\includegraphics[height=5cm,clip=]{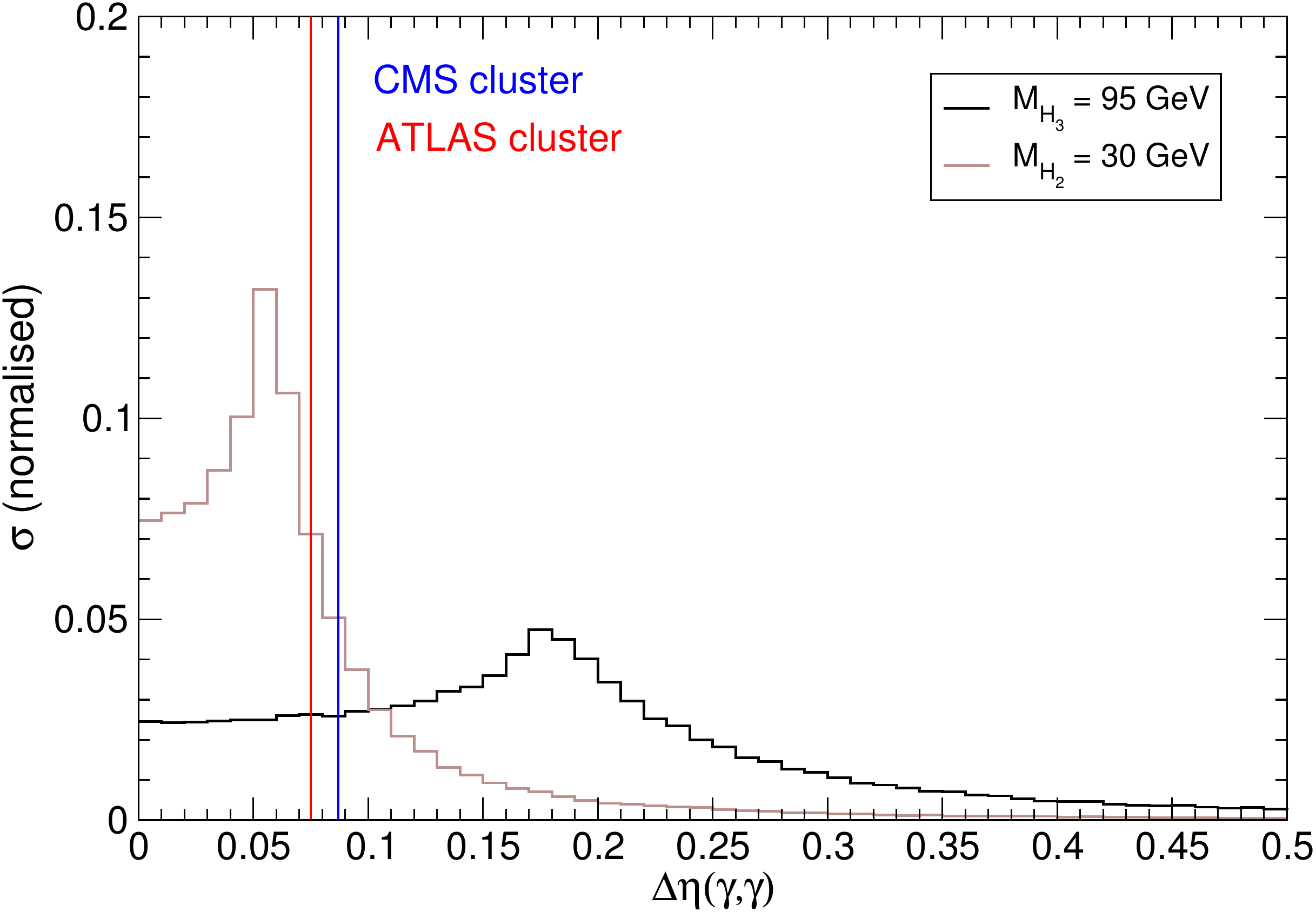}
\end{tabular}
\caption{Parton-level distributions for $Z' \to H_{2,3} X, H_{2,3} \to \gamma \gamma$, being $X$ some additional particle. Left: azimuthal angle difference between the two photons. Right: rapidity separation between the two photons.  For comparison, the size of the ECAL clusters used for photon reconstruction in the CMS and ATLAS detectors is shown with vertical solid lines. The vertical dashed lines indicate the maximum extension in $\Delta \phi$ in case of photon conversion.}
\label{fig:AA}
\end{center}
\end{figure}

Let us now consider the decay $Z' \to H_i X$ with $M_{Z'} = 2.2$ TeV, being $X$ an additional particle (e.g. another scalar of our model with an unspecified decay mode). In figure~\ref{fig:AA} we show the normalised distributions of the diphoton separation in $\Delta \phi$ and $\Delta \eta$ when $H_i$ is either $H_2$ or $H_3$ with $M_{H_3} = 96$ GeV and $M_{H_2} = 30$ GeV, together with the sizes of the ATLAS and CMS clusters.
For $M_{H_2} = 30$ GeV, the two photons lie within the same cluster most of the time. Thus, one expects that an event selection for prompt photons may be sensitive to this decay. 
When $M_{H_3} = 96$~GeV, the two photons are more separate and seldom hit the ECAL within the same cluster. The probability for that to happen can be estimated from the distributions shown in figure~\ref{fig:AA}, being 0.046 for CMS and 0.057 for ATLAS. But the efficiency for the detection of a diphoton as an isolated photon, taking into account the possible extension in $\phi$ of the clusters in case of photon conversion, cannot be simply estimated at the level of fast simulation.

An ATLAS search further attempts to discriminate between a single photon and more than one photon by using the fine granularity of the first layer of the ECAL~\cite{Aaboud:2018djx}. Although this search looks for signatures with a pair of highly-collinear photons, the event selection is only sensitive to $M_{H_i} / M_{Z'} \leq 0.01$, not covering the 30 GeV benchmark mass considered. For example, the limit obtained for $M_{Z'} = 2$ TeV, $M_{H_i,H_j} \lesssim 10$ GeV is $\sigma(pp \to Z') \times \text{BR}(Z' \to H_i H_j \to 4\gamma) \lesssim 0.15$ fb. Therefore, it is safe to say that a dedicated search into pairs of collimated diphotons (or a diphoton plus a jet) using the information from adjacent clusters and applying ad-hoc isolation criteria will be sensitive to the cross sections expected from eq.~(\ref{ec:xsecmax2}).

\subsection{Multiphotons from $H_3 \to H_2 H_2 \to 4\gamma$}
\label{sec:6.2}

The direct production $gg \to H_3 \to H_2 H_2 \to 4 \gamma$ (or, equivalently, $H_4 \to H_2 H_2 \to 4 \gamma$) produces a signal with four well-separated photons. An ATLAS search for events with three or more photons~\cite{Aad:2015bua} in Run 1 at 8 TeV did not observe any significant excess over the SM background expectation, setting an upper limit of 171 fb on the cross section for $gg \to H \to aa \to 4 \gamma$, with $m_a = 10$ GeV. We can conservatively use the same limit for $gg \to H_3 \to H_2 H_2$, bearing in mind that the efficiency will be slightly smaller for $M_{H_3} < M_H$. Using a cross section $\sigma(gg \to h) = 26.5$ pb for a 95 GeV SM-like scalar~\cite{deFlorian:2016spz}, and assuming a 100\% branching ratio of $H_3 \to H_2 H_2$, we can obtain the upper bound
\begin{equation}
|O_{13}| \; \text{BR}(H_2 \to \gamma \gamma) \leq 0.08 \,.
\end{equation}
With our scan over parameter space we have verified that $\text{BR}(H_2 \to \gamma \gamma)$ can be as large as unity fulfilling the above limit.

\begin{figure}[t]
\begin{center}
\begin{tabular}{cc}
\includegraphics[height=5cm,clip=]{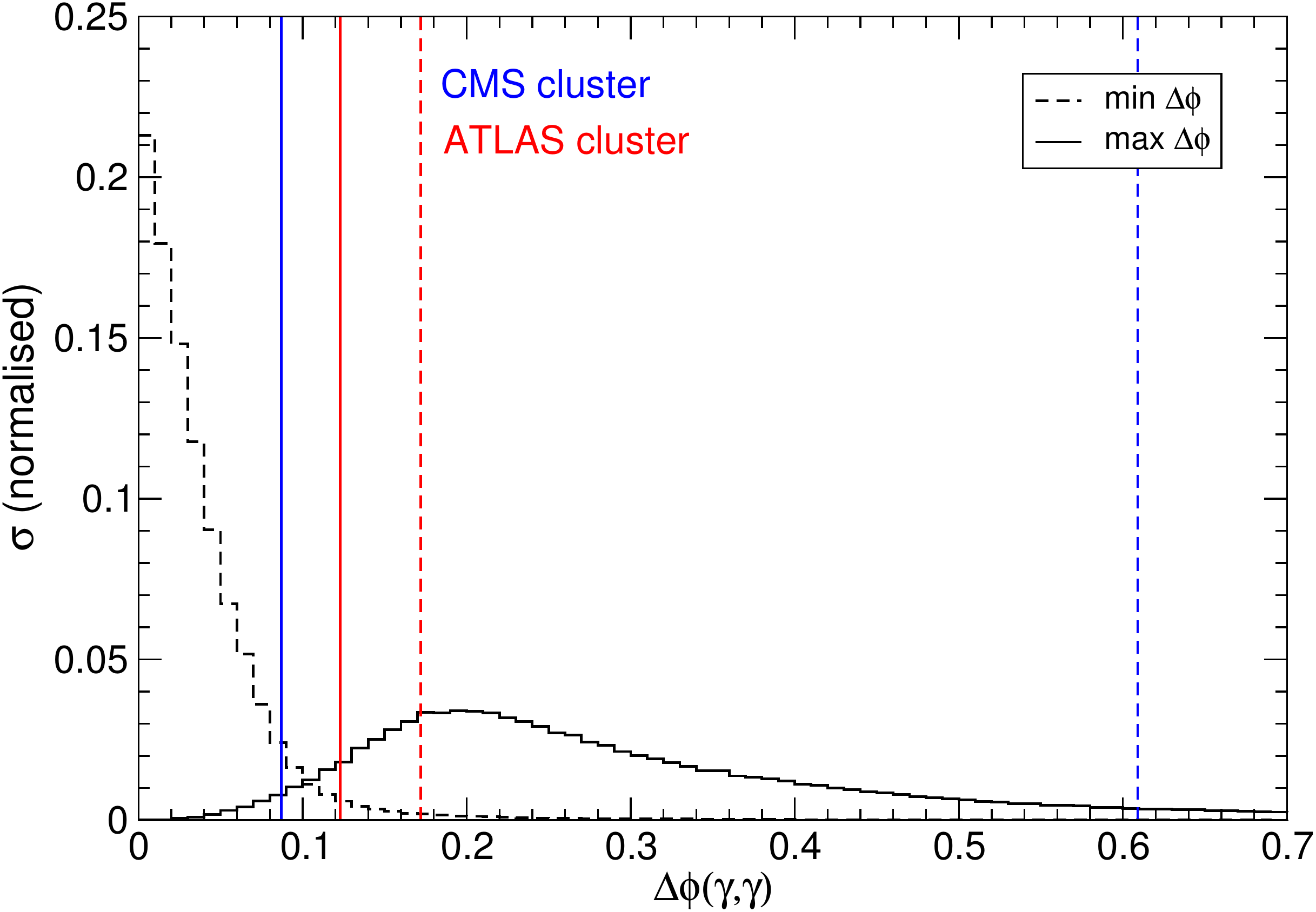} & 
\includegraphics[height=5cm,clip=]{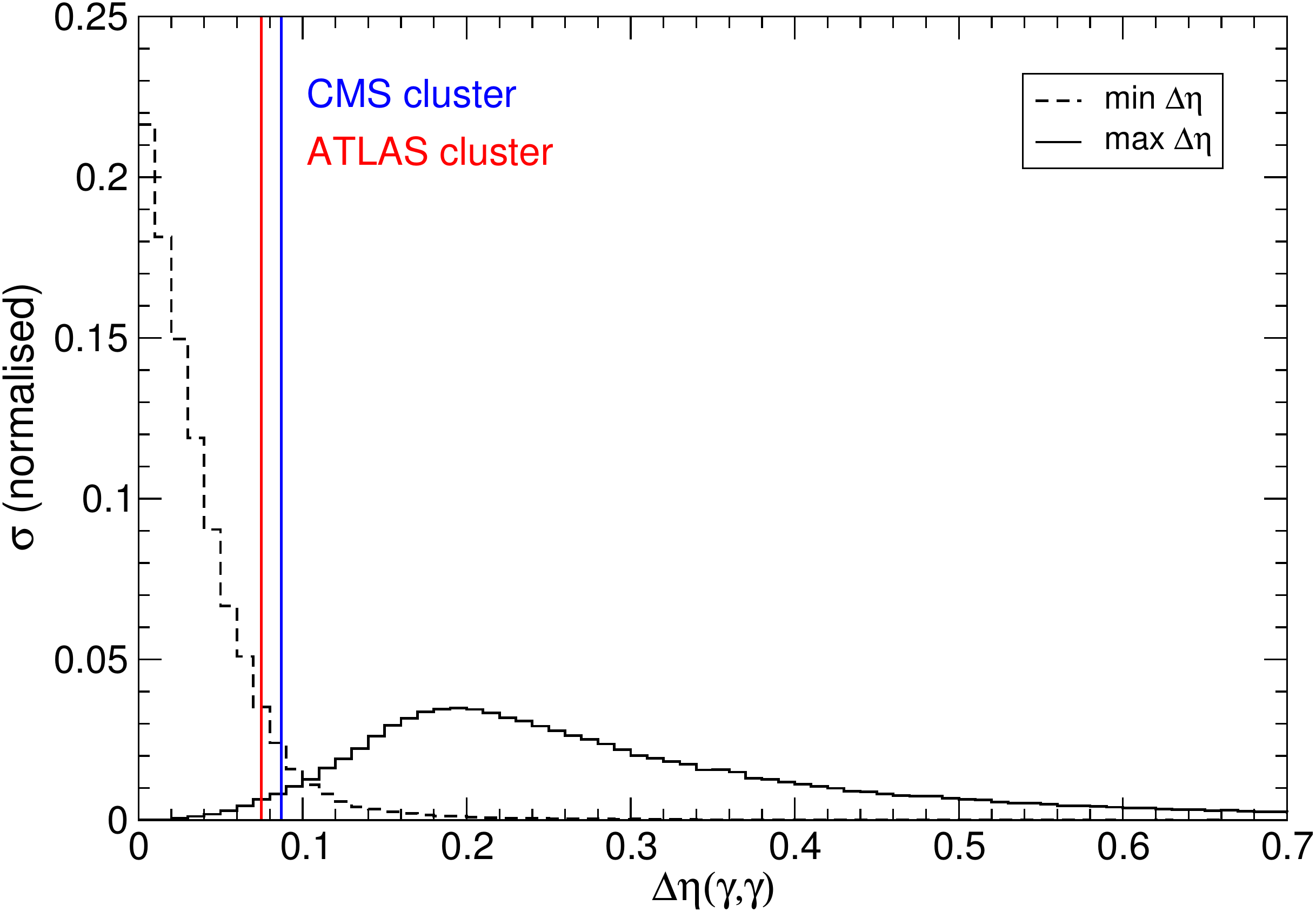}
\end{tabular}
\caption{Parton-level distributions for $Z' \to H_3 X, H_3 \to H_2 H_2 \to 4\gamma $, being $X$ some additional particle. Left: minimum and maximum azimuthal angle difference among the photons. Right: the same, for the rapidity difference. For comparison, the size of the ECAL clusters used for photon reconstruction in the CMS and ATLAS detectors is shown. The dashed lines indicate the maximum extension in $\Delta \phi$ in case of photon conversion.}
\label{fig:4A}
\end{center}
\end{figure}

The cascade decay of a boosted scalar $H_3 \to H_2 H_2 \to 4 \gamma$ yields collinear multiphotons that are much alike the collinear diphotons examined above. We show in figure~\ref{fig:4A} the kinematical distributions for the minimum and maximum separation in $\eta$ and $\phi$ among the four photons, taking $M_{Z'} = 2.2$ TeV, $M_{H_3} = 96$ GeV, $M_{H_2} = 30$ GeV. 
It follows from the previous discussion that some of the photons may be caught within the same cluster, but there are often additional photons that produce energy depositions outside it, thereby spoiling the standard isolation requirements. Such signal also has little or no coverage at all by current searches. We point out that, as in the previous case of boosted diphotons, modified isolation criteria may be applied in dedicated searches for collimated multiphotons, that is, with the radiation contributions to the isolation cone from other photons subtracted before applying the isolation requirements.

\subsection{Photons within jets: $H_i \to H_j H_j \to \gamma \gamma b \bar b$}
\label{sec:6.3}

In analogy with the previous case,  $gg \to H_3 \to H_2 H_2 \to \gamma \gamma b \bar b$ produces a signal with two photons and two $b$ quarks that are generally isolated. 
Searches for heavy scalars $h$ decaying into Higgs boson pairs, $gg \to h \to H H \to \gamma \gamma b \bar b$, are carried out at the LHC~\cite{Sirunyan:2018iwt,Aaboud:2018ftw} but they obviously do not cover the range of interest $M_{H_3} \leq M_H$. On the other hand, searches for exotic decays of the Higgs boson into a pair of lighter particles $H \to aa$ are investigated in a variety of final states, but not with a photon pair. Besides, those analyses are restricted to a narrow mass window around $M_{H}$.

\begin{figure}[htb]
\begin{center}
\begin{tabular}{cc}
\includegraphics[height=5cm,clip=]{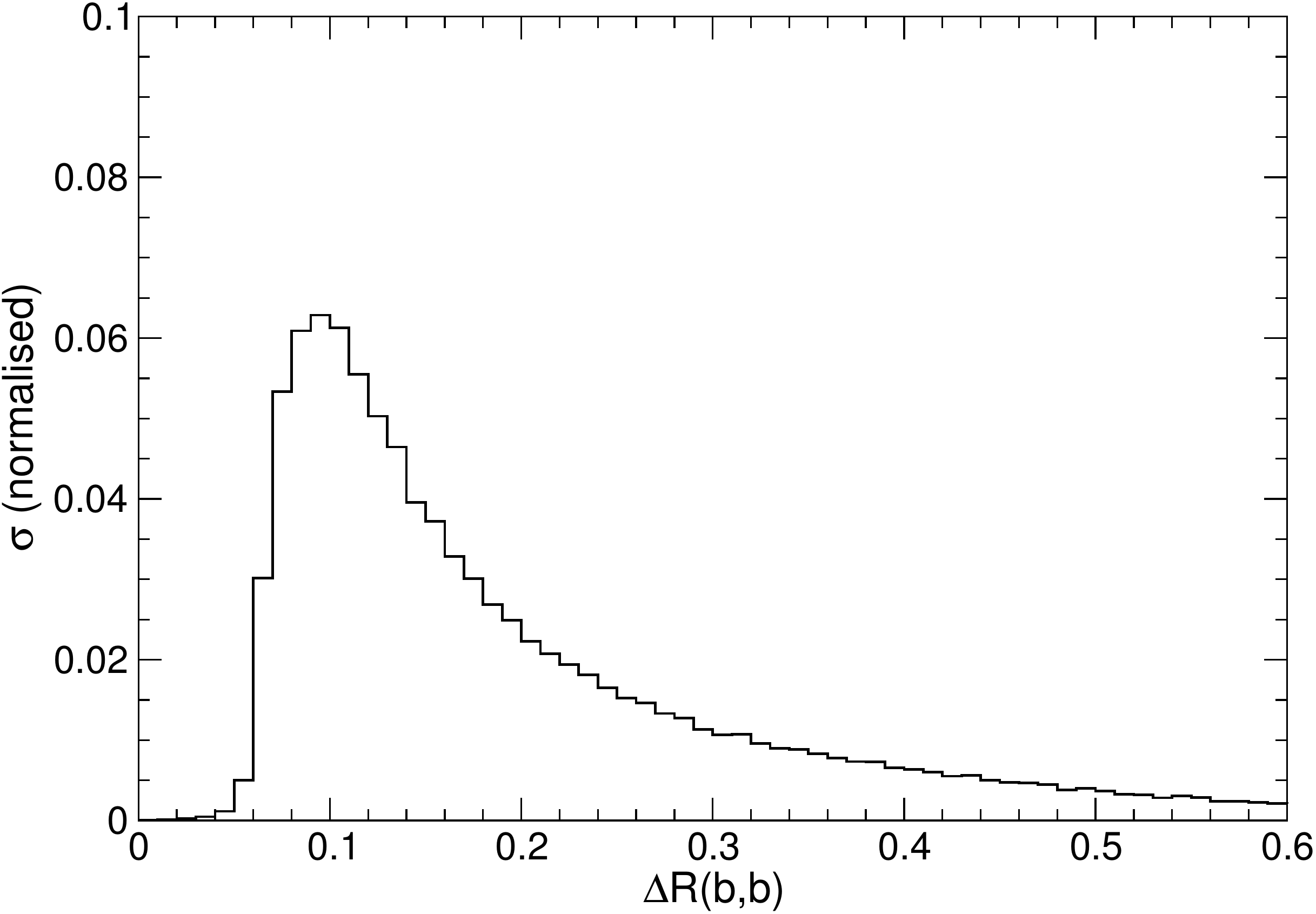} &
\includegraphics[height=5cm,clip=]{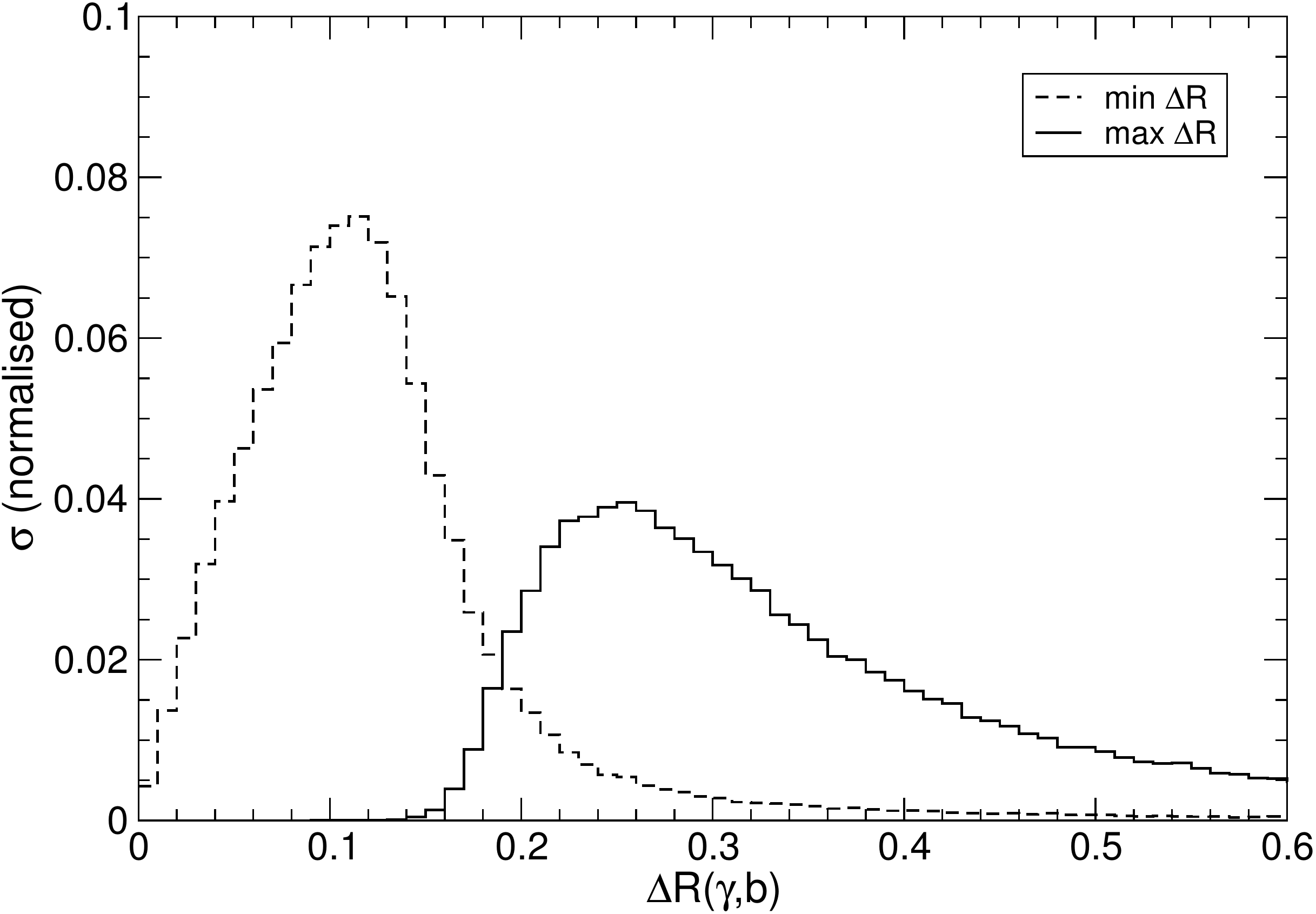}
\end{tabular}
\caption{Parton-level distributions for $Z' \to H_3 X, H_3 \to H_2 H_2 \to \gamma \gamma b \bar b$, being $X$ some additional particle. Left: $\Delta R$ separation between the two $b$ quarks. Right: minimum and maximum $\Delta R$ separation between the photons and $b$ quarks. }
\label{fig:2A2b}
\end{center}
\end{figure}

\begin{figure}[htb]
\begin{center}
\includegraphics[height=5cm,clip=]{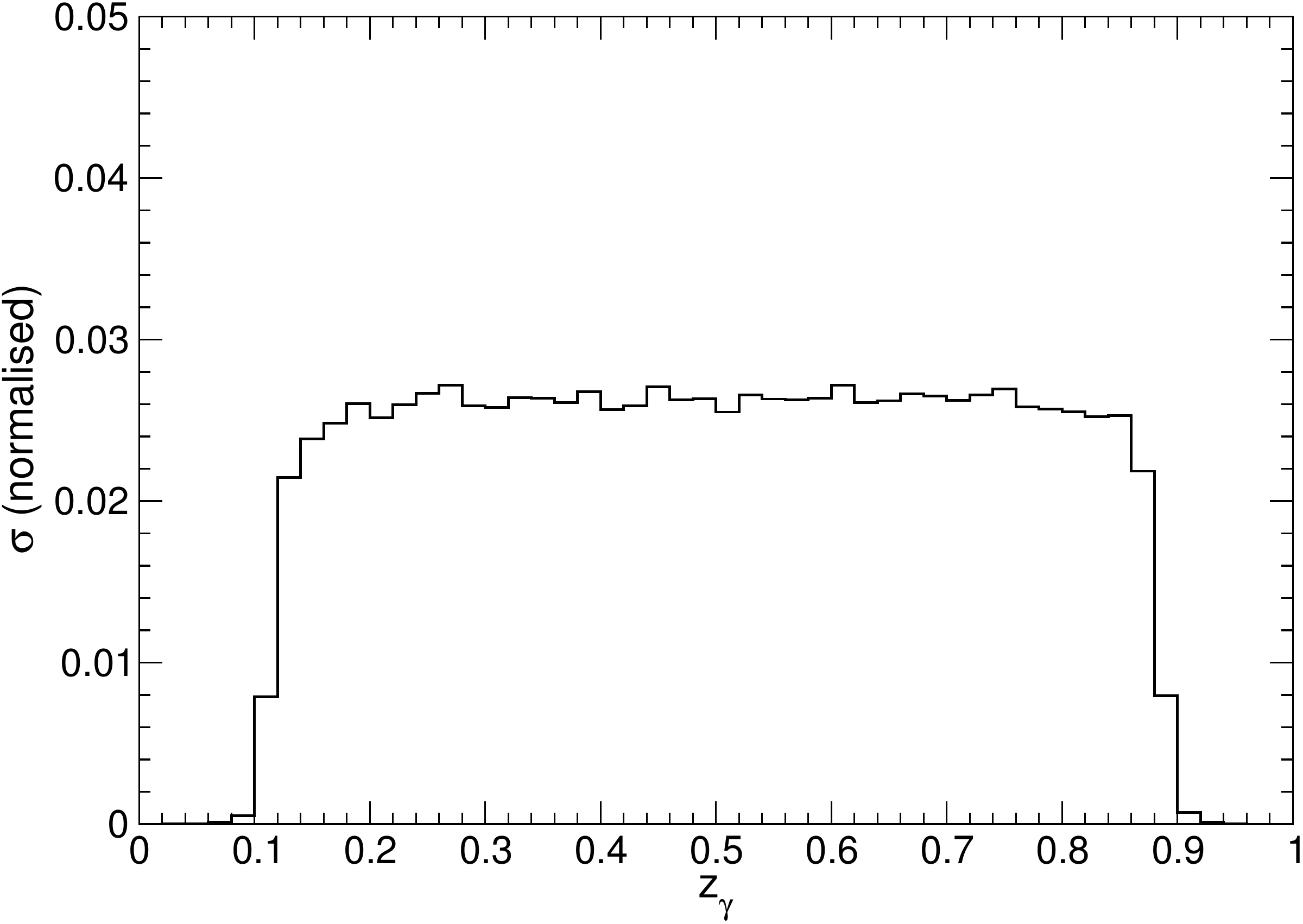}
\caption{Parton-level distributions for the photon energy fraction $z_\gamma$ of jets produced in $Z' \to H_3 X, H_3 \to H_2 H_2 \to \gamma \gamma b \bar b$, being $X$ some additional particle. }
\label{fig:zgamma}
\end{center}
\end{figure}

The decay of a boosted $H_3 \to H_2 H_2 \to \gamma \gamma b \bar b$ produces a conspicuous jet containing hard photons. Such jet is generally narrow, at it can be inferred from the distribution of the $\Delta R$ distance between the two $b$ quarks in figure~\ref{fig:2A2b} (left). The two photons are 
at a medium distance from the $b$ quarks, as shown in the right panel. A quite distinctive feature of these jets is the large energy fraction carried by the photons. We can define it as
\begin{equation}
z_\gamma = \frac{E_{\gamma_1} + E_{\gamma_2}}{E_{b_1} + E_{b_2}} \,,
\end{equation}
where the subindices 1 and 2 label the different photons and $b$ quarks, and the energies are considered in the laboratory frame. The kinematical distribution of this quantity, depicted in figure~\ref{fig:zgamma}, shows that a large fraction of the jet energy may be carried by the photons. As antecipated above, this is a quite peculiar signature. Searches for such objects seem difficult but feasible, indeed the CMS Collaboration has searched for events with jets that contain one photon and two gluons~\cite{Sirunyan:2019ape}.

\section{The 96 GeV case}
\label{sec:5}

The diphoton signals from direct production of new scalars bear a special interest in view of some CMS measurements which we now summarise. The CMS Collaboration has found~\cite{Sirunyan:2018aui} two nearby excesses when searching for new scalars $h$ decaying into photon pairs in Run 1 and Run 2 data,
\begin{itemize}
\item Run 1: $2\sigma$ local significance at a diphoton invariant mass $m_{\gamma \gamma} = 97.6$ GeV, with 19.7 fb$^{-1}$ at 8 TeV;
\item Run 2: $3\sigma$ local significance at a slightly lower mass $m_{\gamma \gamma} = 95.3$ GeV, with 35.9 fb$^{-1}$ at 13 TeV.
\end{itemize}
They also provide limits on two groups of processes mediating the production of $h$: (a) $gg \to h$ plus $t \bar th$ production; (b) VBF plus $Vh$ production. The excess is present in both, though more prominent in the former group. The reported combination of Run 1 and Run 2 data has a maximum $2.8\sigma$ local significance at $m_{\gamma \gamma} = 95.3$ GeV. This significance is slightly smaller than that of Run 2 data alone because the two excesses are not located precisely at the same mass, and the mass shift is larger than the detector resolution. However, one cannot discard the possibility that the slight mass difference arises from some statistical or systematic effect: one can also notice a shift {\em in the same direction} between the peaks at $M_Z$, coming from $Z \to e^+ e^-$, with the electrons misidentified as photons. In Run 1, the peak is located at 91.0 GeV, while in Run 2 it is at 90.0 GeV. In this regard, it is also worth remembering that the final combined Run 1 measurements of the Higgs boson mass by the ATLAS Collaboration in the cleanest decay modes $H \to \gamma\gamma$ and $H \to ZZ^*$ yielded masses $M_H = 125.98 \pm 0.50$ GeV and $M_H = 124.51 \pm 0.51$ GeV, respectively, with uncertainties dominated by statistics~\cite{Aad:2014aba}. These two measurements were compatible only at the $2\sigma$ level even if the collected samples were large enough to have $5.2\sigma$ and $8.1\sigma$ significances for the Higgs signal, respectively, in the $\gamma \gamma$ and $ZZ^*$ channel.

The CMS excess corresponds to a signal strength $\mu_\text{CMS} = 0.6 \pm 0.2$~\cite{Biekotter:2019kde}, defined as
\begin{equation}
\mu_\text{CMS} 
= \frac{\sigma(gg \to h \to \gamma \gamma)}{\sigma(gg \to h \to \gamma \gamma)_\text{SM}}
= \frac{\sigma(gg \to h)}{\sigma(gg \to h)_\text{SM}} \times \frac{\text{BR}(h \to \gamma \gamma)}{\text{BR}(h \to \gamma \gamma)_\text{SM}} \,,
\label{ec:mucms}
\end{equation}
with the subindex `SM' indicating that the quantity is referred to a SM-like scalar $h$, of course evaluated for its corresponding mass $M_h$. This excess can be accommodated with an eigenstate that is mainly a singlet, by either enhancing the production via gluon fusion or the decay into $\gamma \gamma$. As the CMS excess is also seen in the VBF and $Vh$ category, the latter possibility seems quite reasonable. Since in our model the new particles do not contribute to the $gg \to H_i$ amplitudes, then the cross-section ratio in the r.h.s. of eq.~(\ref{ec:mucms}) reduces to the mixing factor $O_{13}^2$. (As mentioned, we consider $H_3$ as being the new scalar candidate to explain the excess.) Taking $\text{BR}(H_3 \to \gamma \gamma)_\text{SM} = 1.43 \times 10^{-3}$~\cite{deFlorian:2016spz}, we find
\begin{equation}
O_{13}^2 \, \text{BR}(H_3 \to \gamma \gamma) = (8.6 \pm 2.9) \times 10^{-4} \,.
\label{ec:hgam}
\end{equation}
From figure~\ref{fig:Hdecays} it is clear that the agreement with (\ref{ec:hgam}) is possible even for small $O_{13}$ mixing, and in the presence of non-SM decay modes. Also, we see that it is quite possible that one scalar gives an observable signal $pp \to H_3 \to \gamma \gamma$ while the others do not, if the mixings $O_{12}$ and $O_{14}$ are smaller.

\begin{figure}[t!]
\begin{center}
\includegraphics[height=6cm,clip=]{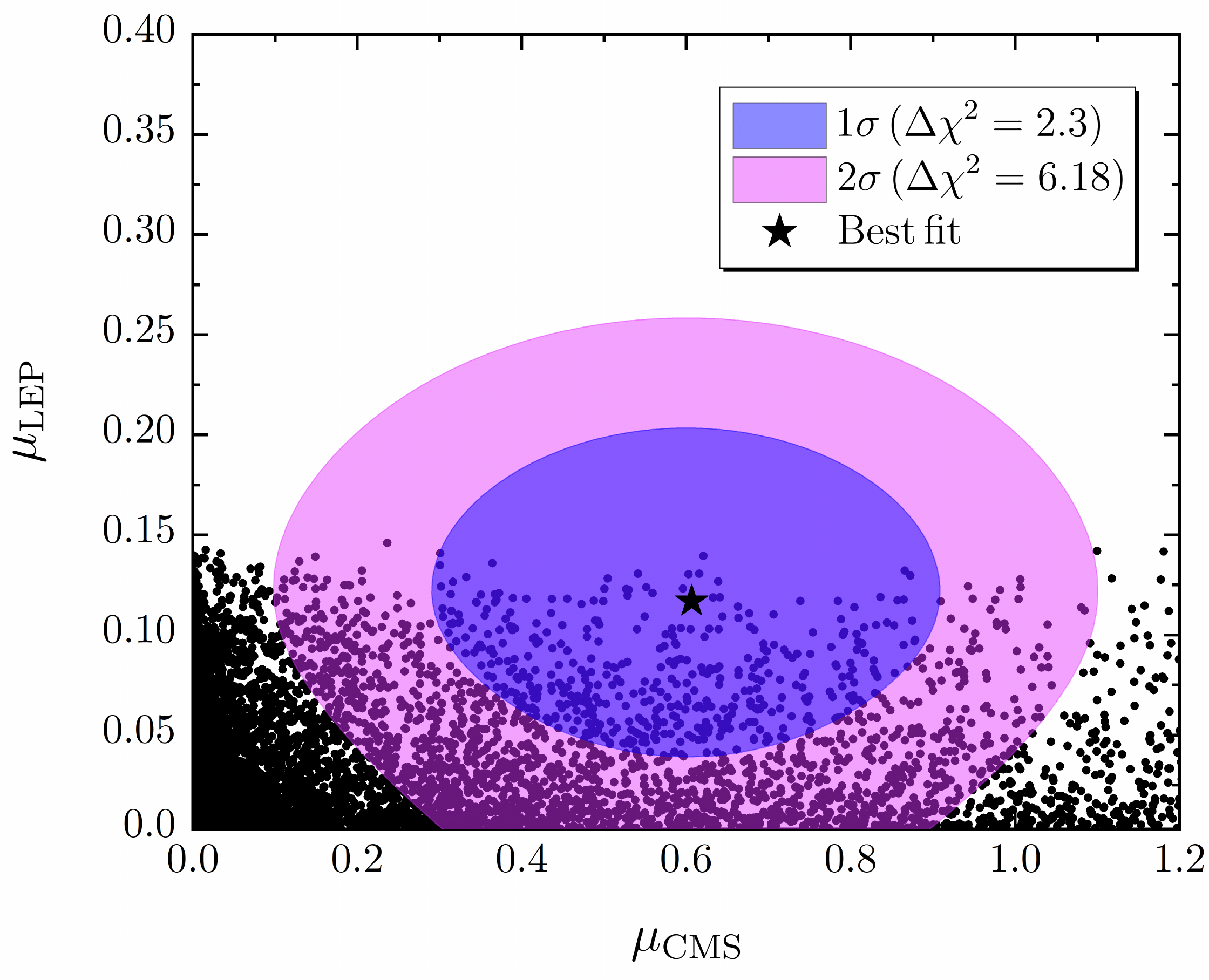}
\caption{Allowed range for $\mu_\text{CMS}$ and $\mu_\text{LEP}$, as defined in eqs. (\ref{ec:mucms}), (\ref{ec:mulep}) obtained from the parameter-space scan. The best-fit point corresponds to $\mu_\text{CMS} = 0.6 \pm 0.2$~\cite{Biekotter:2019kde}, $\mu_\text{LEP} = 0.117 \pm 0.057$~\cite{Cao:2016uwt}.}
\label{fig:mus}
\end{center}
\end{figure}

The CMS excess has sparked some interest in the framework of SM extensions with a scalar singlet~\cite{Fox:2017uwr,Biekotter:2017xmf,Hollik:2018yek,Domingo:2018uim,Biekotter:2019kde,Cline:2019okt,Choi:2019yrv,Liu:2018xsw,LiuLiJia:2019kye,Biekotter:2019gtq,Cao:2019ofo}, also in connection with a $2.3\sigma$ local excess observed at a mass of 98 GeV in LEP searches for the SM Higgs boson ($e^+ e^- \to Zh$, $h \to b \bar{b} $)~\cite{Barate:2003sz}. The fact that the two excesses are located at nearly the same mass, suggests that they could originate from $\gamma\gamma$ and $b \bar{b}$ decays of the same particle. The signal strength of the LEP excess, defined as
\begin{equation}
\mu_\text{LEP} 
= \frac{\sigma(e^+ e^- \to Z h \to Z b \bar b )}{\sigma(e^+ e^- \to Z h \to Z b \bar b)_\text{SM}}
= \frac{\sigma(e^+ e^- \to Z h)}{\sigma(e^+ e^- \to Z h)_\text{SM}} \times \frac{\text{BR}(h \to b \bar b)}{\text{BR}(h \to b \bar b)_\text{SM}} \,,
\label{ec:mulep}
\end{equation}
has been computed in ref.~\cite{Cao:2016uwt}, yielding $\mu_\text{LEP} = 0.117 \pm 0.057$. The subindex `SM' refers to quantities corresponding to a SM-like scalar with mass $M_h$. In our case, the cross section ratio in the r.h.s. of eq.~(\ref{ec:mulep}) is again $O_{13}^2$. Taking $\text{BR}(h \to b \bar b)_\text{SM} = 0.8$~\cite{deFlorian:2016spz}, eq.~(\ref{ec:mulep}) implies
\begin{equation}
O_{13}^2 \, \text{BR}(h \to b \bar b) = 0.0934 \pm 0.045 \,,
\label{ec:constLEP}
\end{equation}
As shown in figure~\ref{fig:mus}, both the CMS and LEP excesses can be accommodated in the MSBM. The value of $\mu_\text{CMS}$ can be either larger or smaller than the best-fit value, in agreement with the results in figure~\ref{fig:Hdecays}. On the other hand, $\mu_\text{LEP}$ cannot be much larger than the best-fit value, as it is limited by the Higgs constraints on $O_{13}^2$ --- see eq.~(\ref{ec:constLEP}). Note that in our benchmark scenario, the decay $H_3 \to H_2 H_2$ is allowed; for heavier $H_2$ this channel would be closed and the lower left side of the plot with both $\mu_\text{CMS}$ and $\mu_\text{LEP}$ small would be less populated.

\begin{figure}[t!]
\begin{center}
\begin{tabular}{cc} \\
\includegraphics[height=6cm,clip=]{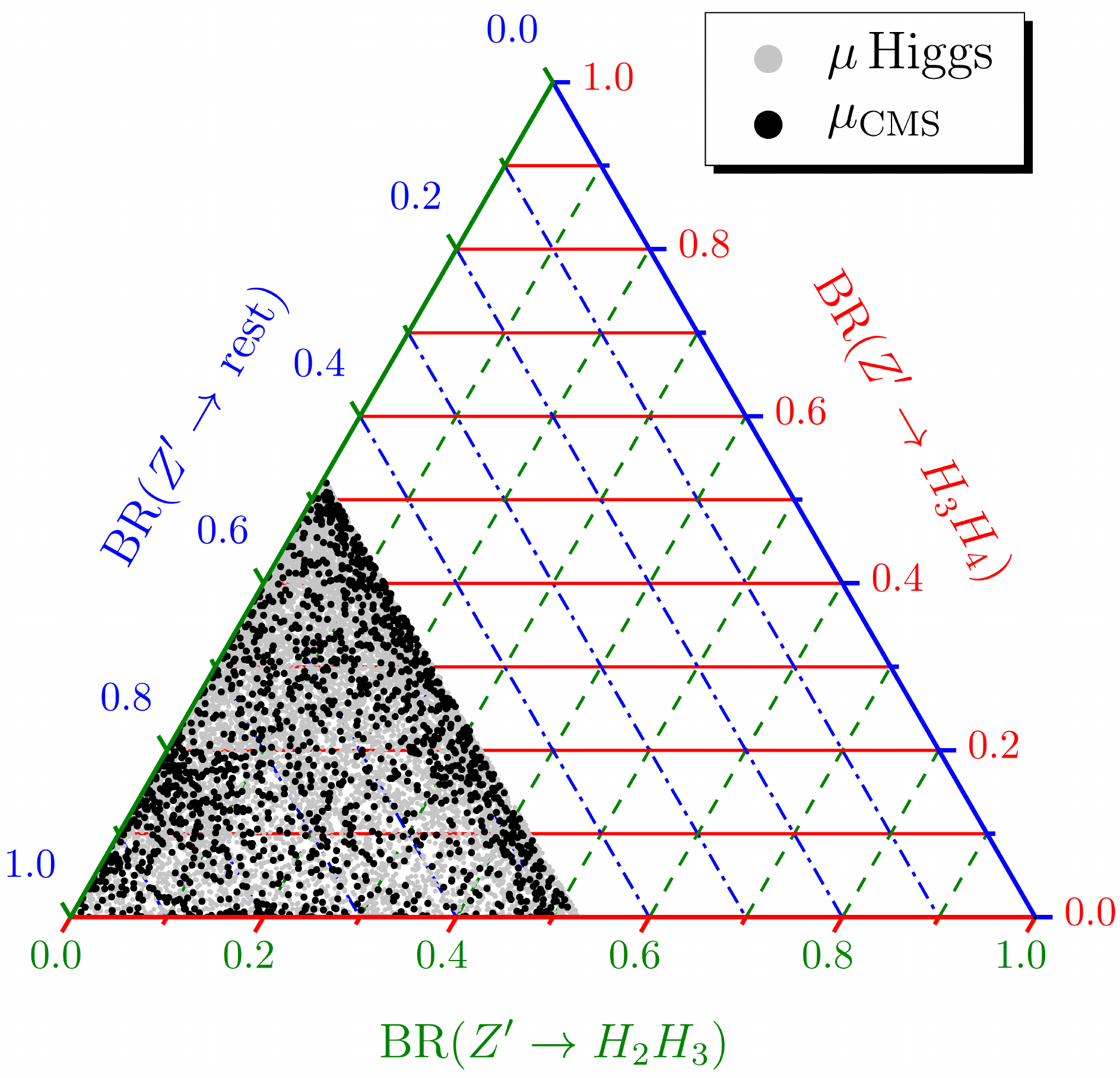} & 
\includegraphics[height=6cm,clip=]{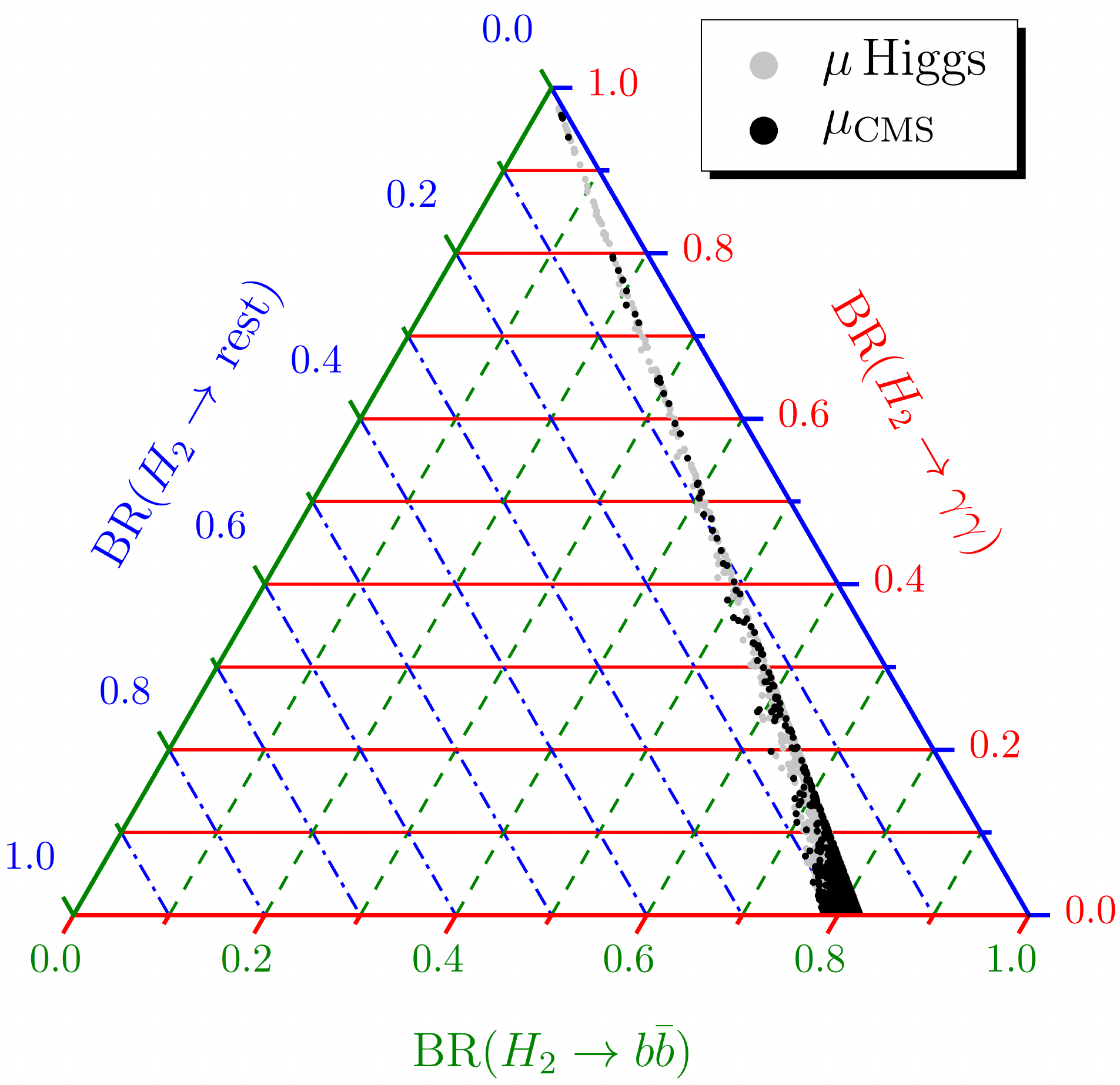} \\[2mm]
\includegraphics[height=6cm,clip=]{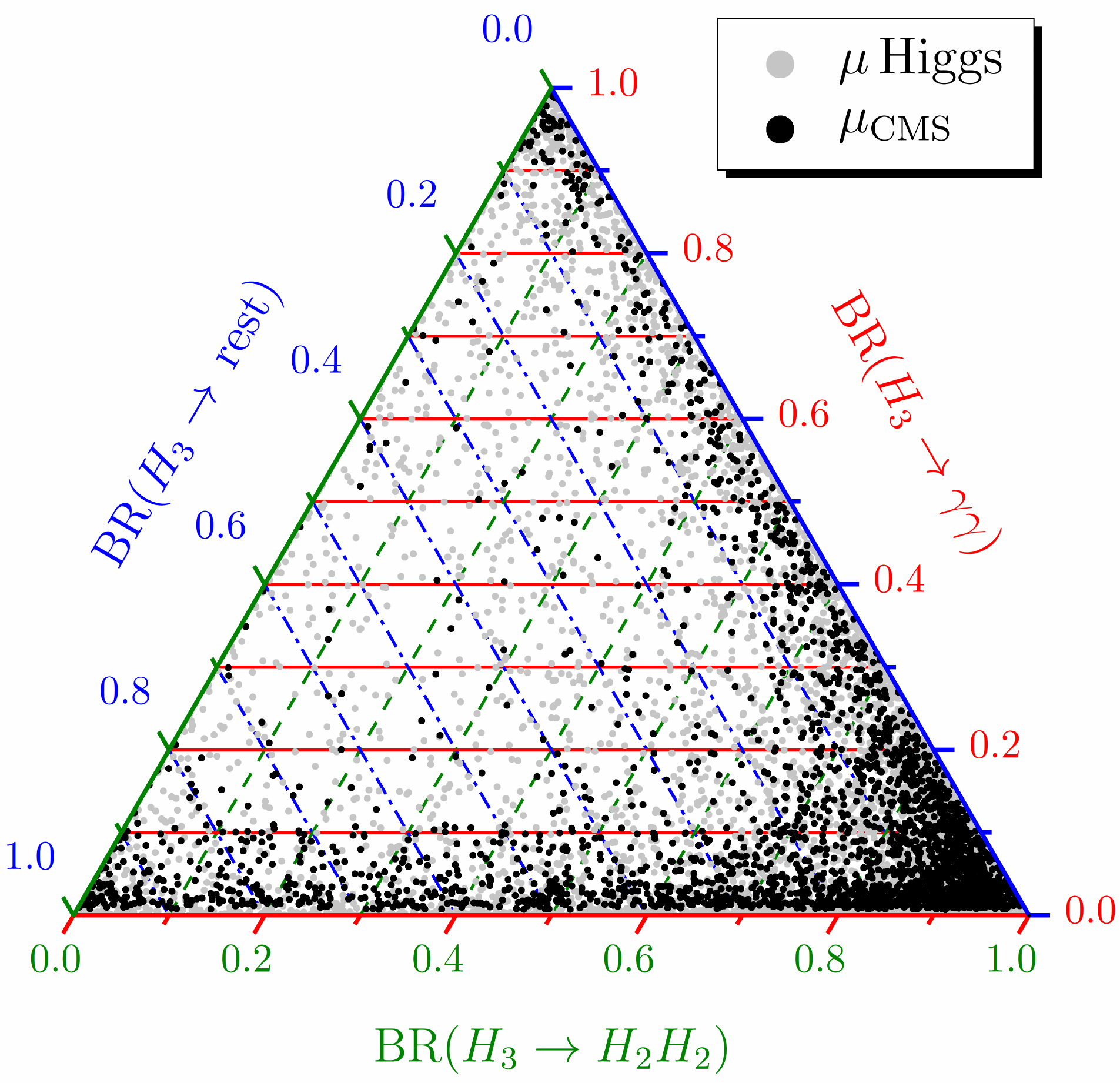} & 
\includegraphics[height=6cm,clip=]{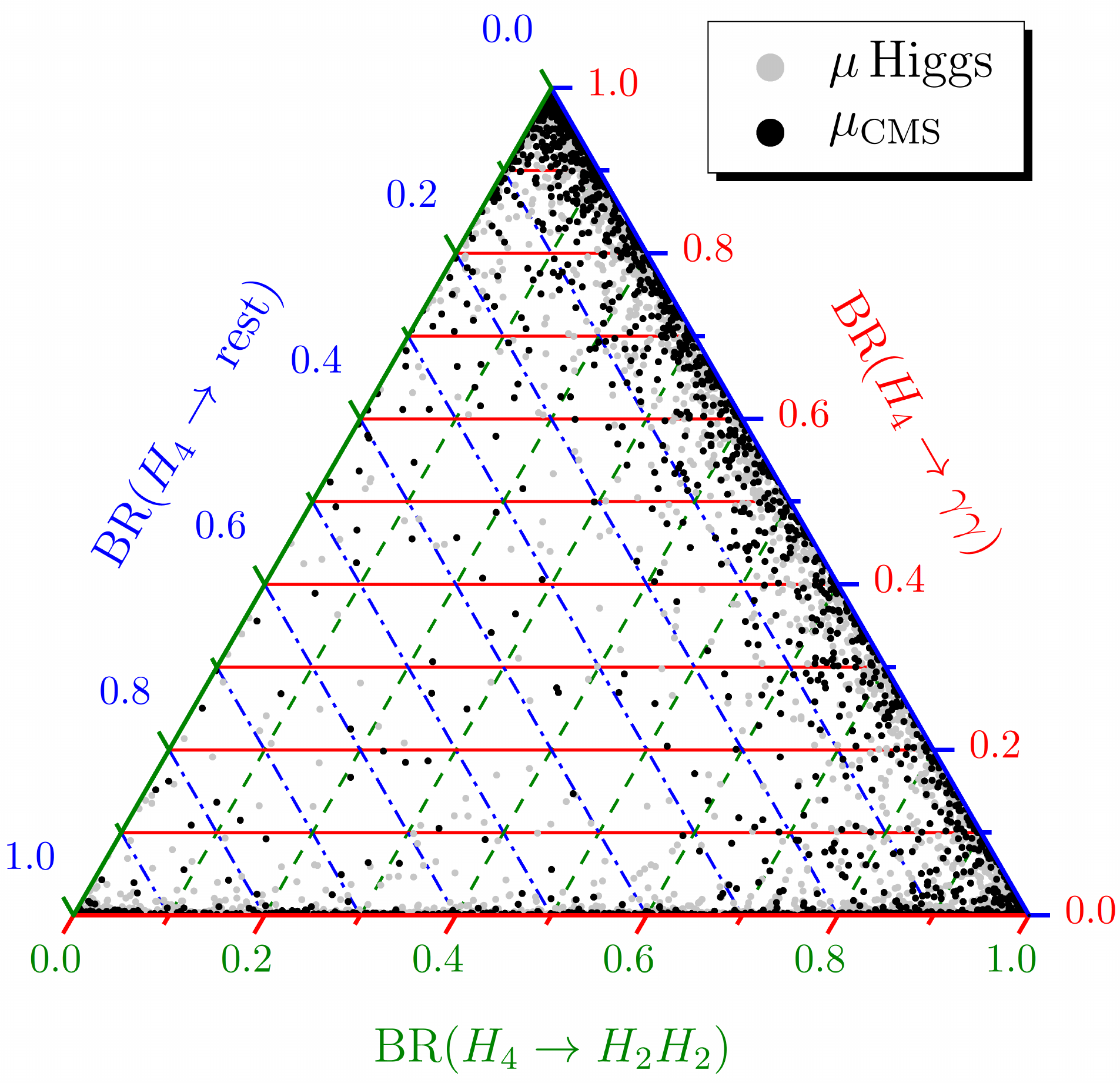}
\end{tabular}
\caption{Branching ratios for the selected $Z'$ and $H_i$ decays. While all points (gray and black) obey the constraints on the Higgs signal-strength parameters in eqs.~\eqref{eq:muggval} and \eqref{eq:murestval}, only the black ones are compatible with the CMS excess.}
\label{fig:ZpHdecays96}
\end{center}
\end{figure}

The scenarios we are interested in are those in which $H_3 \to H_2 H_2$ (as well as $H_4 \to H_2 H_2$) have sizeable branching ratios, thus producing the multiphoton signals described in the previous section. Therefore, we drop the interpretation of the LEP excess, which requires large branching ratio $H_3 \to b \bar b$ (see the appendix). 
When we require $\mu_\text{CMS} = 0.6 \pm 0.2$, the phenomenology remains mostly the same even if the parameter space is reduced. This is shown in figure~\ref{fig:ZpHdecays96}. On the top-left panel we observe that the pattern of $Z'$ decays into scalars remains practically unaltered. The same happens for the lightest scalar $H_2$ (top right). For $H_3$ (bottom left) there is a stronger preference for $H_3 \to H_2 H_2$ decays, and for $H_4$ (bottom right) the behaviour is similar. The gray points in these plots correspond to the points in figures~\ref{fig:Zpdecays} and \ref{fig:Hdecays}, while the black points are those which are also compatible with the CMS excess, with $\mu_\text{CMS} \in [0.4,0.8]$. Therefore, one can conclude that the multiphoton signals studied in section~\ref{sec:4} are compatible with the CMS excess too.

\section{Summary and discussion}
\label{sec:6}

New scalars are thoroughly searched for at the LHC. Still, there is a number of theoretically and phenomenologically well-motivated signatures that are yet to be explored, even for scalars lighter than the Higgs boson. The measurement of the couplings of the SM Higgs~\cite{Aad:2019mbh} suggests that new scalars, if they exist, have small couplings to the weak bosons.\footnote{This is for example the case of two Higgs doublet models, where in the alignment limit the new neutral scalars $H^0$, $A^0$ do not couple to $WW$ and $ZZ$, and of models with additional singlets, where the coupling of the mass eigenstate is acquired via a small doublet component.} Therefore, provided these new scalars have small Yukawa coupling, their most visible signals may arise when produced from the decay of a new heavy resonance. (In case of very small mixing, these could even be the signals with larger cross section.) On the other hand, the cascade decay into lighter scalars may be the dominant decay mode, if kinematically allowed. In this context, we have investigated a number of collider signals involving diphotons from heavy resonance decays into scalars.

We have worked within the framework of the MSBM~\cite{Aguilar-Saavedra:2019adu}, an anomaly-free leptophobic $\text{U}(1)_{Y'}$ extension of the SM, in which anomaly cancellation is ensured by a set of vector-like leptons. Additionally to the SM Higgs doublet, two complex singlets charged under $\text{U}(1)_{Y'}$ are introduced with the purpose of providing mass to the $Z'$ boson and the new leptons, and of allowing for $Z'$ decays into scalars. Overall, the model contains four physical scalar particles $H_i$, being $H_1 \equiv H$ the 125 GeV Higgs boson. To provide an interpretation for a CMS diphoton excess, we choose the mass of $H_3$ to be 96 GeV. Moreover, a lighter scalar ($H_2$) with mass in the $20-40$ GeV range has also been considered to allow for the decay $H_3 \to H_2 H_2$. Since the mass of $H_4$ does not play a significant role in the signals involving $H_3$ and $H_2$, we take $M_{H_4}=105$~GeV (relatively close to $M_{H_3}$). Its signals are analogous to the ones with $H_3$.

We have performed a scan over the MSBM parameter space in order to explore the allowed range of the mixing and the decay modes of the new scalars, compatible with existing constraints from several SM Higgs production and decay processes~\cite{Aad:2019mbh}. We have found that the allowed $\text{BR}(H_{2-4} \to \gamma \gamma)$ span over the whole range from zero to unity, while keeping $\text{BR}(H \to \gamma \gamma)$ for the SM Higgs boson in agreement with current measurements. This is possible due to the fact that the scalar singlets couple to the new leptons (they actually provide their mass), while the SM Higgs doublet does not. Thus, the $H_{2-4} \to \gamma \gamma$ loop amplitudes can be enhanced, while those for $H \to \gamma \gamma$ and, of course, $gg \to H$ are not modified.

Direct production of $H_{2-4}$ is suppressed by the small mixing with the SM Higgs doublet. Still, we have found that a CMS excess at 96 GeV may be easily accommodated by one of the $H_i$ (which we have labelled as $H_3$) since, as mentioned, the decay width to $\gamma \gamma$ can be enhanced. If the dominant decay mode is $H_3 \to b \bar b$, then the same scalar can also explain a LEP excess found around the same mass. However, the most intriguing consequences arise when the main decay channel of this particle is $H_3 \to H_2 H_2$, which is otherwise the most natural situation if there is a lighter scalar $H_2$ and available phase space. From figure~\ref{fig:ZpHdecays96}, we observe that in most of the parameter space where $H_3 \to \gamma \gamma$ fits the CMS excess:
\begin{itemize}
\item[(i)] $H_3$ predominantly decays into $H_2 H_2$, with $\text{BR}(H_3 \to \gamma \gamma)$ at the percent level. (Note however that there are points where $H_3 \to \gamma \gamma$ dominates.)
\item[(ii)] $H_2 \to b \bar b$ is expected to dominate.
\item[(iii)] The main decay modes of $H_4$ are $H_2 H_2$ and $\gamma \gamma$.
\end{itemize}
We can then have $pp \to Z' \to H_3 H_4$, with $H_3 \to \gamma \gamma$ and $H_4 \to H_2 H_2 \to 4b$. (If, instead, we consider the decay $H_3 \to H_2 H_2 \to 4b$, the signal contains two four-pronged jets.) This signal yields a pair of collinear photons from $H_3$, plus a multi-pronged jet from $H_4$ which is quite elusive and requires specific tools to be efficiently separated from the QCD background (see for example refs.~\cite{Aguilar-Saavedra:2017rzt,Heimel:2018mkt,Farina:2018fyg}). So does the boosted diphoton which, as seen in section~\ref{sec:6.1}, will usually fail the isolation criteria for prompt photons and requires a specific selection. The decay $pp \to Z' \to H_3 H_2$ produces a collinear photon pair plus a two-pronged jet of very small mass, which is hard to separate from the QCD background.

It is worth recalling that the CMS Collaboration has reported another excess when searching for heavy resonances decaying into $Z\gamma$, with 35.9 fb$^{-1}$ of data at 13 TeV~\cite{Sirunyan:2017hsb}. The excess reaches a local significance of $3.6\sigma$ at a mass of 2 TeV in the hadronic channel, without any excess in the leptonic channel. In a search with similar sensitivity~\cite{Aaboud:2018fgi}, the ATLAS Collaboration did not find any excess at that mass. The aforementioned process $pp \to Z' \to H_{3} H_4$, with $H_{3} \to \gamma \gamma$ and $H_4 \to H_2 H_2 \to 4b$, is an obvious candidate to explain an excess in the hadronic channel without a counterpart in the leptonic channel. (Note that the pruned jet mass distribution of the $H_4$ jet with $M_{H_4} = 105$ GeV falls within the window selected in the analysis, as the pruning significantly decreases the mass of multi-pronged jets~\cite{Aguilar-Saavedra:2018xpl}.) However, this interpretation faces two difficulties. First and foremost, the fact that the CMS Collaboration does {\em not} see any excess in an event category with $b$-tagged jets. A multi-pronged jet with four $b$ quarks should in principle yield an excess in this category, although the $b$ tagging efficiency for such a jet can only be verified with a full simulation of the CMS detector. Second, the fact that the excess is not seen in the ATLAS analysis does not have an obvious explanation: as we have mentioned in section~\ref{sec:6.1}, the efficiency for a boosted diphoton to be identified as a photon depends on details of the photon reconstruction that are not available without the full detector simulation. Besides, with a fast simulation we have estimated that the efficiency for the fat jet from $H_4 \to 4b$ to pass the jet substructure selection in the ATLAS and CMS analyses is similar.

Dropping the connection with the CMS 96 GeV excess, the cascade decay of the heavier scalars with diphoton decay of the lighter one may give two interesting signals:
\begin{itemize}
\item[(i)] $gg \to H_3 \to H_2 H_2 \to 4 \gamma$. This signal is covered by a search for events with at least three photons by the ATLAS Collaboration using 8 TeV data~\cite{Aad:2015bua}. To the best of our knowledge, there are no searches at 13 TeV in this final state.
\item[(ii)]  $gg \to H_3 \to H_2 H_2 \to \gamma \gamma b \bar b$. There are searches for heavy scalars decaying into the Higgs boson in this final state~\cite{Sirunyan:2018iwt,Aaboud:2018ftw} but for lighter $H_3$ this is not covered.
\end{itemize}
Note that the same signals can appear with the scalar $H_4$ replacing $H_3$, but for brevity we only refer to the latter.
The fully hadronic signals $H_3 \to H_2 H_2 \to 4b$ have quite large backgrounds and small sensitivity. 
In $WH/ZH$ associated production, a search for $H \to aa \to b \bar b b \bar b $ (where $a$ is a lighter particle, e.g. an axion) by the ATLAS Collaboration~\cite{Aaboud:2018iil} has a sensitivity one order of magnitude smaller than would be required, and does not cover $M_{H_3} = 96$ GeV. At the Tevatron, the CDF Collaboration performed a search for pair production of new particles $Y$, each decaying into two jets, $p \bar p \to YY \to jjjj$~\cite{Aaltonen:2013hya}. The mass range explored $M_{Y} \geq 50$ GeV does not cover $M_{H_2} = 30$ GeV and, in addition, the sensitivity is two orders of magnitude below the possible size of the signals.

The same cascade decays, when $H_3$ is produced in the decay of the $Z'$ boson, yield boosted and collinear final state particles that form complex objects:
\begin{itemize}
\item[(i)] $Z' \to H_3 X \to 4 \gamma + X$, with $X$ an additional particle. The four photons are produced closely (see figure~\ref{fig:4A}) and require a special object selection.
\item[(ii)] $Z' \to H_3 X \to \gamma \gamma b \bar b+ X$. The two $b$ quarks are produced close in $\Delta R$ and merge into a single jet; the same jet contains two very energetic photons. The identification of such an object requires a specific tagging.
\end{itemize}
This type of final-state objects (collinear multiphotons and jets with photons) are not currently searched for. Notice that the `photon jets' searched for by the ATLAS Collaboration~\cite{Aaboud:2018djx} are much more collinear, and that search does not cover the multiphotons produced at intermediate distances, such as those shown in figure~\ref{fig:4A}.
 
To conclude, cascade decays including multiphotons in the final state yield intriguing signals that are not covered by current experimental searches. The CMS 96 GeV excess, whether corresponds to a new particle or not, provides an extra motivation to explore these final states. In direct connection with this excess, a dedicated search for heavy resonances decaying into collimated diphotons plus a fat (multi-pronged) jet would be of great interest. Especially, because it might clarify the situation regarding the 2 TeV excess in the search for heavy resonances decaying into $Z\gamma$.

\section*{Acknowledgements} J.A.A.S. thanks J. Alcaraz, A. Casas, M.C. Fouz, C. Mu\~noz and J. Terr\'on for useful discussions. F.R.J. is grateful for the warm hospitality and financial support from IFT/UAM-CSIC (Madrid) where most of this work was done. He also acknowledges  Funda\c{c}{\~a}o para a Ci{\^e}ncia e a Tecnologia (FCT, Portugal) for support through the projects CFTP-FCT Unit 777 (UID/FIS/00777/2013 and UID/FIS/00777/2019), CERN/FIS-PAR/0004/2017 and PTDC/FIS-PAR/29436/2017, which are partly funded through POCTI (FEDER), COMPETE, QREN and EU. The work of J.A.A.S. is supported by Spanish Agencia Estatal de Investigaci\'on through the grant `IFT Centro de Excelencia Severo Ochoa SEV-2016-0597' and by MINECO project FPA 2013-47836-C3-2-P (including ERDF).

\appendix
\section{Scalar decays and interpretation of the CMS and LEP excesses}
\label{sec:a}

\begin{figure}[htb]
\begin{center}
\begin{tabular}{cc} \\
\multicolumn{2}{c}{\includegraphics[height=6cm,clip=]{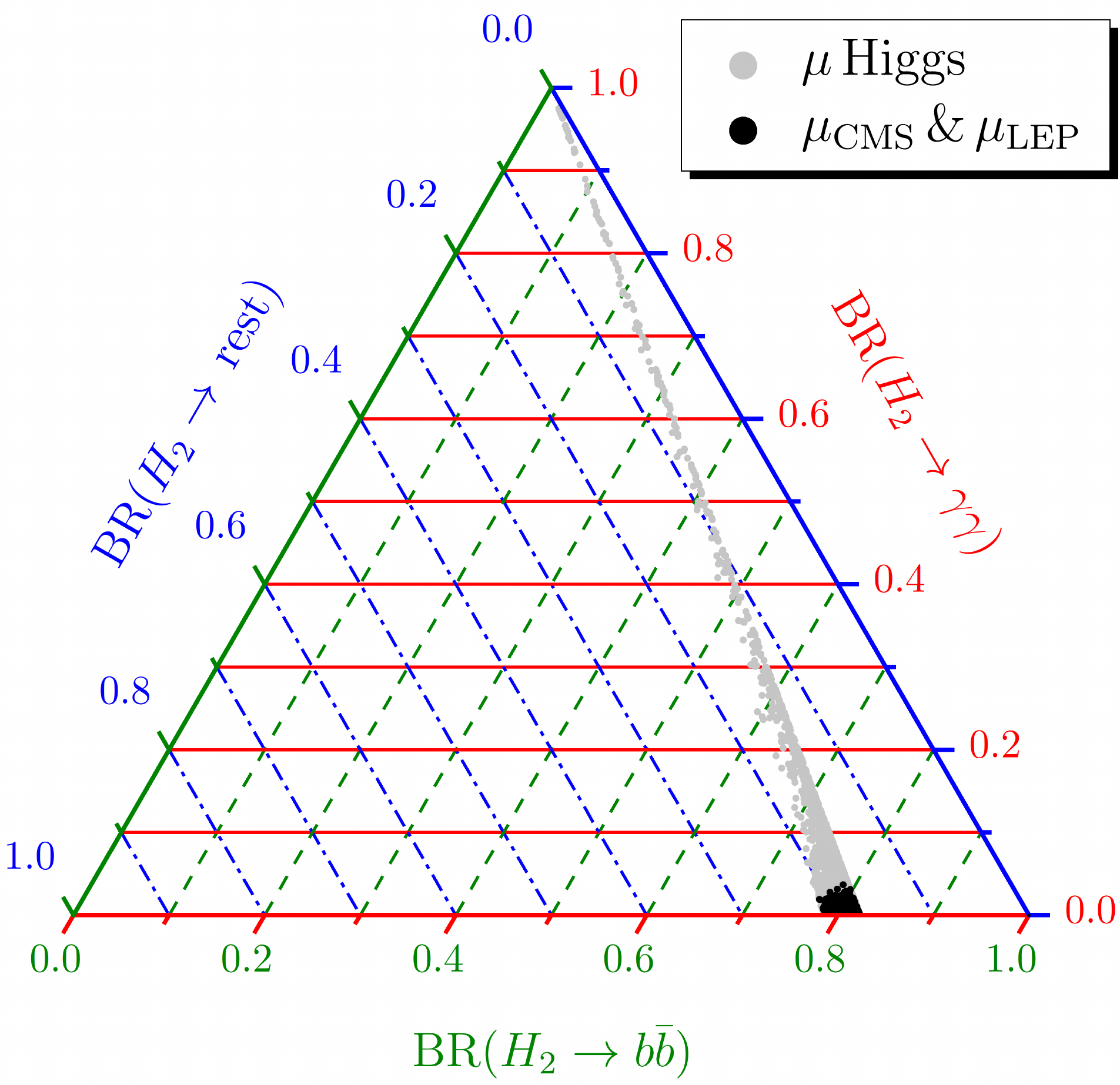}} \\[2mm]
\includegraphics[height=6cm,clip=]{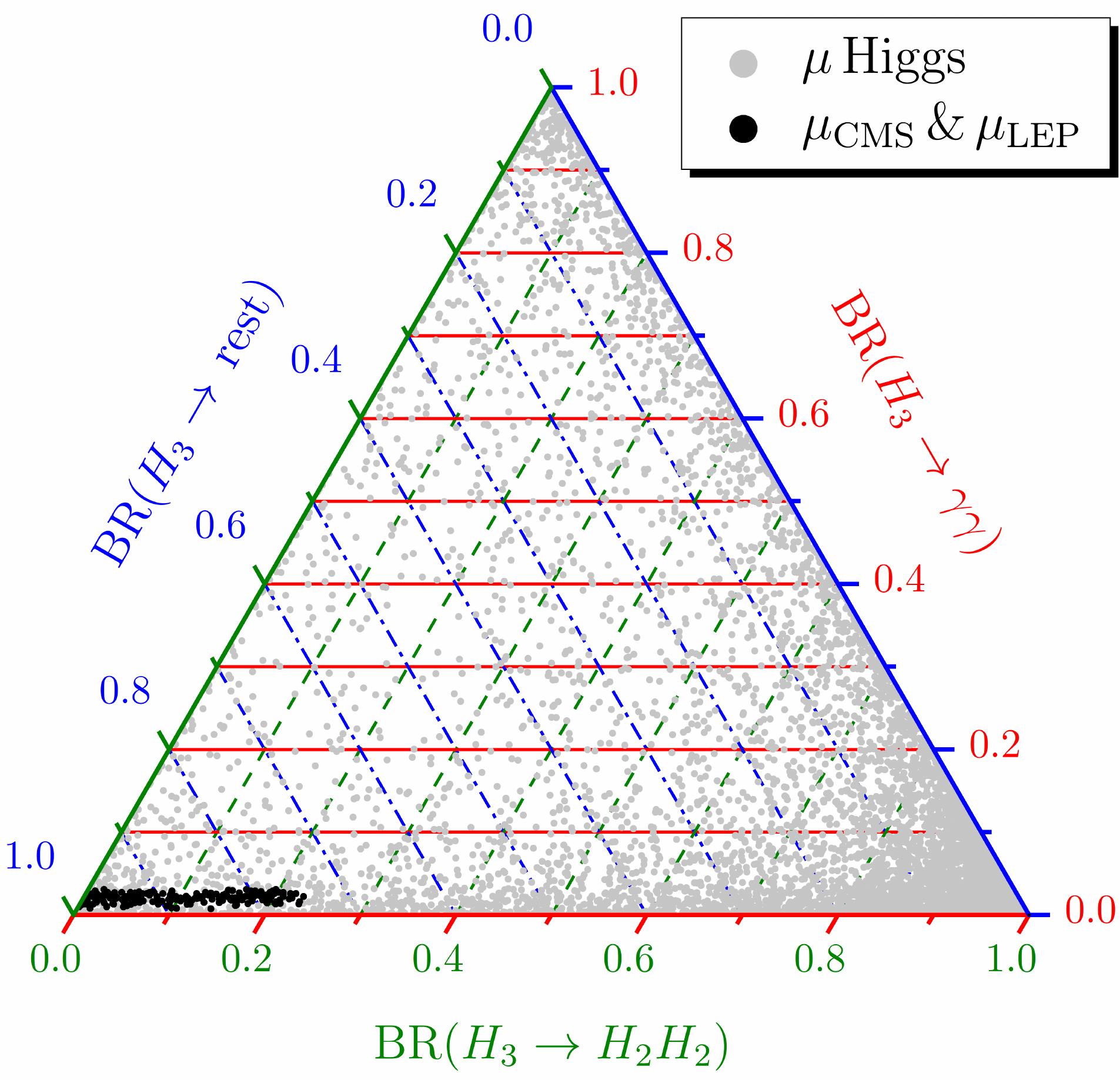} & 
\includegraphics[height=6cm,clip=]{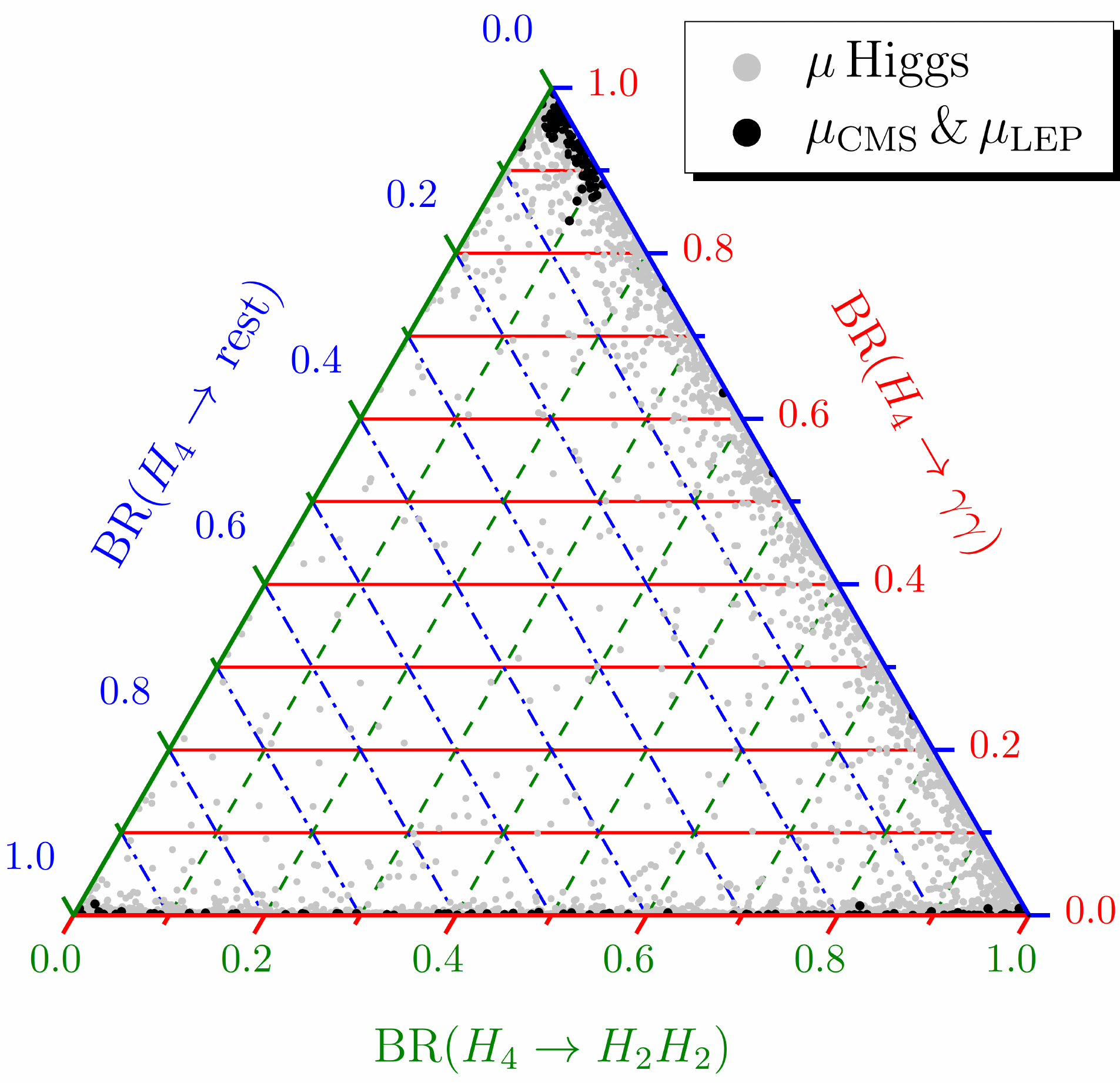}
\end{tabular}
\caption{Decay BRs for the new scalars into selected final states. All points (gray and black) obey the constraints on the Higgs signal-strength parameters in eqs.~\eqref{eq:muggval} and \eqref{eq:murestval}. Black points are also compatible with the CMS and LEP excesses.}
\label{fig:Hdecays96LEP}
\end{center}
\end{figure}

As seen in figure~\ref{fig:mus}, the CMS and LEP excesses can be simultaneously fitted within our model. We present in figure~\ref{fig:Hdecays96LEP} the allowed BRs of the three scalars in such case. Besides the fact that $\text{BR}(H_3 \to b \bar b)$ has to be large, as it can easily inferred from eq.~(\ref{ec:constLEP}), the lightest scalar does not decay into $\gamma \gamma$ and, on the other hand, $\text{BR}(H_4 \to \gamma \gamma)$ is close to unity. Therefore, as anticipated, the cascade decays $H_{3,4} \to H_2 H_2$ that produce the signals studied in section~\ref{sec:4} have small BRs.

\end{document}